\documentclass[]{aastex631}

\usepackage{amsmath}

\shorttitle{Nonlinear Alfv\'en Waves with Viscosity Tensor}
\shortauthors{Russell}

\graphicspath{{./}{figures/}}

\begin{document}

\title{Nonlinear Damping and Field-Aligned Flows of Propagating Shear Alfv\'en Waves with Braginskii Viscosity}
\author[0000-0001-5690-2351]{Alexander J. B. Russell}
\affiliation{
School of Science and Engineering, \\
University of Dundee, \\
Dundee, DD1 4HN, Scotland, UK}
\correspondingauthor{Alexander J. B. Russell}
\email{a.u.russell@dundee.ac.uk}

\begin{abstract}
Braginskii MHD provides a more accurate description of many plasma environments than classical MHD since it actively treats the stress tensor using a closure derived from physical principles. Stress tensor effects nonetheless remain relatively unexplored for solar MHD phenomena, especially in nonlinear regimes. This paper analytically examines nonlinear damping and longitudinal flows of propagating shear Alfv\'en waves. Most previous studies of MHD waves in Braginskii MHD considered the strict linear limit of vanishing wave perturbations. We show that those former linear results only apply to Alfv\'en wave amplitudes in the corona that are so small as to be of little interest, typically a wave energy less than $10^{-11}$ times the energy of the background magnetic field. For observed wave amplitudes, the Braginskii viscous dissipation of coronal Alfv\'en waves is nonlinear and a factor around $10^9$ stronger than predicted by the linear theory. Furthermore, the dominant damping occurs through the parallel viscosity coefficient $\eta_0$, rather than the perpendicular viscosity coefficient $\eta_2$ in the linearized solution. This paper develops the nonlinear theory, showing that the wave energy density decays with an envelope $(1+z/L_d)^{-1}$. The damping length $L_d$ exhibits an optimal damping solution, beyond which greater viscosity leads to lower dissipation as the viscous forces self-organise the longitudinal flow to suppress damping.  Although the nonlinear damping greatly exceeds the linear damping, it remains negligible for many coronal applications. 
\end{abstract}


\section{Introduction}\label{sec:intro}

Alfv\'enic waves are a ubiquitous feature of natural plasmas, including the solar corona \citep{2007Tomczyk,2007DePontieu,2007Lin,2007Okamoto} and solar wind \citep{1967Coleman,1971BelcherDavis}. In solar physics, these waves contain sufficient energy to heat the open corona and accelerate the fast solar wind \citep{2011McIntosh}, and they damp significantly within a solar radius above the surface \citep{2012BemporadAbbo,2012Hahn,2013HahnSavin,2022Hahn}. How these Alfv\'enic waves damp in astrophysical and space plasmas is an important question that has remained open for almost a century (see early papers by \citealt{1947Alfven} and \citealt{1961Osterbrock}; modern reviews by \citealt{2015DeMoortelBrowning}, \citealt{2015Arregui} and \citealt{2020VanDoorsselaere}; and historical perspectives by \citealt{2018Russell} and \citealt{2020DeMoortel}). 

Most theoretical knowledge about solar Alfv\'enic waves is based on ``classical'' magnetohydrodynamics (MHD), a mathematical framework that originated from intuitive coupling of Maxwell's equations and Euler equations of inviscid hydrodynamics \citep{1937Hartmann,1942Alfven,1943Alfven,1950Alfven,1950Batchelor} and became widely adopted in large part due to its success providing insight into diverse natural phenomena \citep[see e.g.][]{2014Priest}. However, classical MHD is one member of a larger family of plasma descriptions, some of which offer a more complete description of the plasma. This paper analytically examines Alfv\'en wave damping in the more general framework of Braginskii MHD, which unlike classical MHD, retains the anisotropic viscous stress tensor.

\begin{figure}
\centering
\includegraphics[width=0.75\columnwidth]{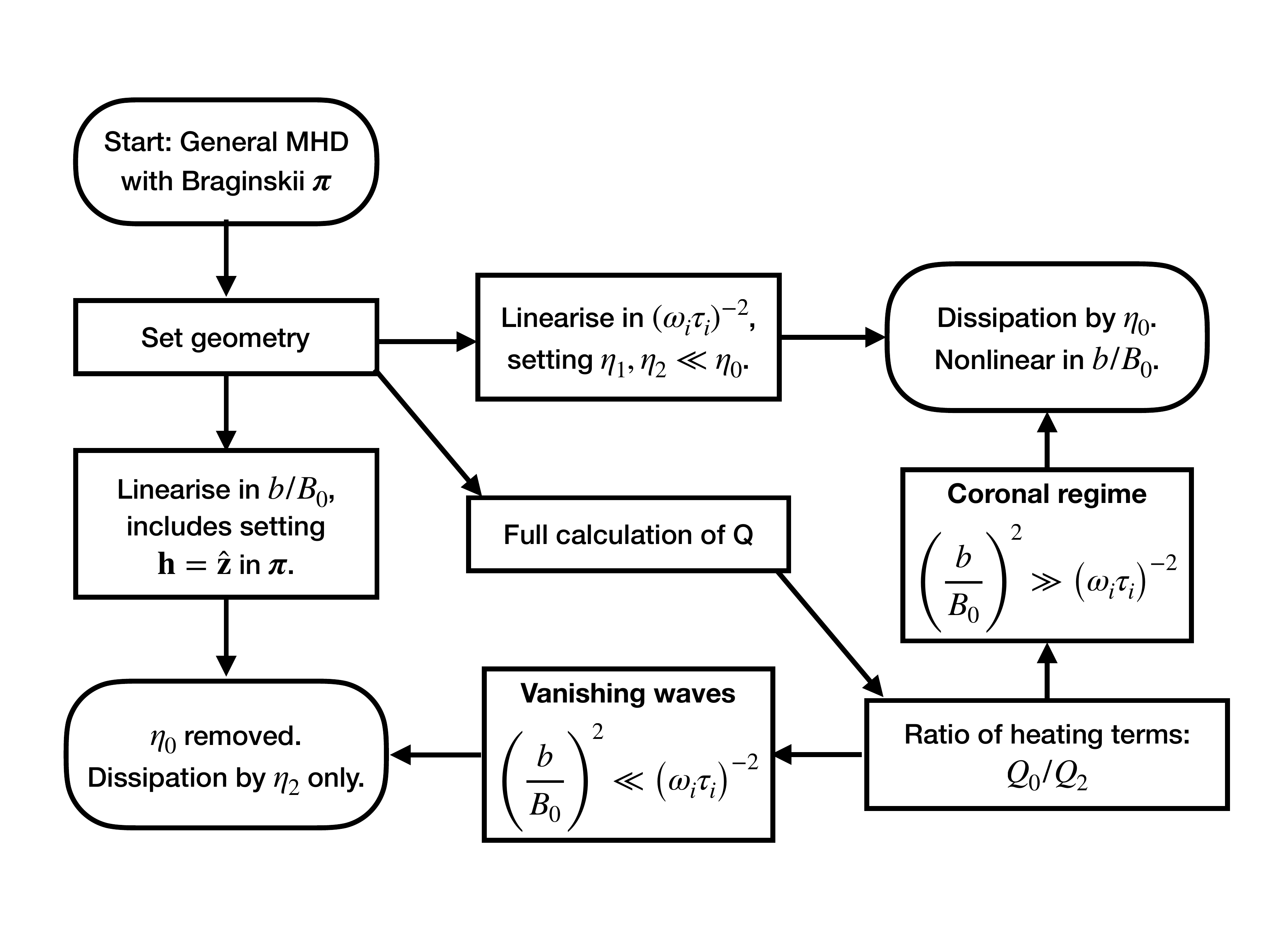}
\caption{Schematic paths of reasoning. The vertical branch gives priority to smallness of the wave amplitude and concludes that damping is a linear process governed by the perpendicular viscosity coefficient $\eta_2$, e.g. \S~8 of \citet{1965Braginskii}. The horizontal branch gives priority to the smallness of $\eta_2/\eta_0$, leading to nonlinear damping via $\eta_0$. This paper follows the diagonal branch, which includes deriving the validity condition for the two outcomes. Nonlinear damping via $\eta_0$ is appropriate for most coronal applications.
}\label{fig:reasoning_paths}
\end{figure}

A number of authors, including \S~8 of \citet{1965Braginskii}, have previously investigated viscous damping of Alfv\'en waves in the linear limit of vanishingly small wave amplitude. When priority is given to smallness of the wave amplitude, the problem becomes framed as a matter of how anisotropic viscosity affects velocities that are perpendicular to the magnetic field (the direction of which is treated as unchanging). With this approximation, damping is determined by the ``perpendicular'' viscosity coefficient $\eta_2$, which is extremely small in the corona. It was thus originally concluded that viscous damping is very weak for coronal Alfv\'en waves unless they have very short wavelengths. This path of reasoning is shown as the vertical branch in Fig.~\ref{fig:reasoning_paths}.

There is, however, another way to view the problem. Viscous damping of Alfv\'en waves can alternatively be considered with priority given to the largeness of parallel viscosity coefficient $\eta_0$. Given that $\eta_2/\eta_0\gtrsim10^{-11}$ is typical in the corona \citep[][]{1985Hollweg}, even a very small component of $\mathbf{v}$ parallel to the total magnetic field $\mathbf{B}$ would be expected to produce major departures from linear theory. This path of reasoning is shown as the horizontal branch in Fig.~\ref{fig:reasoning_paths}.

The second viewpoint of the problem takes impetus from the observation that (unless wave amplitudes vanish entirely) Alfv\'en waves do have a non-zero velocity component parallel to the \textit{total} magnetic field. Two effects contribute to this, which are separated if one expands $\mathbf{V}\cdot\mathbf{B} = \mathbf{V}\cdot(\mathbf{b}+\mathbf{B}_0)$, where $\mathbf{B}_0$ is the equilibrium magnetic field and $\mathbf{b}$ is the magnetic perturbation. First,  $\mathbf{V}\cdot\mathbf{b}$ is non-zero for an Alfv\'en wave, since the velocity perturbation perpendicular to $\mathbf{B}_0$ is aligned with the magnetic field perturbation $\mathbf{b}$. In other words, deflection of the magnetic field from its equilibrium direction implies there is a non-zero velocity component parallel to the \textit{total} magnetic field. Second, in compressible plasma, the magnetic pressure of the magnetic field perturbation drives a nonlinear ponderomotive flow parallel to the equilibrium magnetic field \citep[e.g.][]{1971Hollweg}. The ponderomotive flow makes $\mathbf{V}\cdot\mathbf{B}_0$ nonzero as well. Both of these effects allow for the possibility of nonlinear viscous damping via the large parallel viscosity coefficient $\eta_0$.

With the benefit of modern observations \citep[e.g.][]{2011McIntosh,2015Morton}, it is known that normalised wave amplitudes $b/B_0 \sim V/v_A \sim 0.1$ are typical for the base of an open coronal field region, for example. The ``smallness'' of the square of this ratio is very modest in comparison to the extreme largeness of $\eta_0/\eta_2$. Thus, it is likely from the outset that viscous damping of Alfv\'en waves will be a nonlinear process governed by $\eta_0$ and the wave amplitude. This paper provides mathematical evidence that this heuristic analysis holds true, along with detailed examination of the consequences. 

Various previous studies have explored effects of Braginskii viscosity on MHD waves since \citet{1965Braginskii}. In solar physics, the effect of linearized Braginskii viscosity was revisited from the 1980s to the mid-1990s through the lens of phase mixing and resonant absorption, with the aim of determining how including the viscosity tensor modifies these scale-shortening processes and their heating properties. At the time, it was common practice in solar MHD wave theory to work with linearized equations. Thus, due to linearization, \citet{1986Steinolfson}, \citet{1987Hollweg}, \citet{1991Ruderman}, \citet{1994Ofman} and \citet{1995Erdelyi} obtained analytical and numerical results that strictly apply to Alfv\'en waves of vanishing amplitude.

In adjacent fields, the effect of anisotropic viscosity on MHD waves has also been investigated with an eye on MHD turbulence and the solar wind. Of particular note, \citet{1992Montgomery} advocated that Braginskii viscosity is important in hot tenuous plasmas, that in many circumstances it should be treated using parallel ion viscosity, and that plasma motions may self-organise to suppress damping. He further applied these ideas to anisotropy in MHD turbulence, on the basis that a quasi-steady turbulence is composed of the undamped modes. Quantitative elaboration in \citet{1992Montgomery} was based on a linear normal mode analysis, which captures linear damping of magnetoacoustic waves by parallel viscosity, but excludes nonlinear viscous damping of Alfv\'en waves. The conclusion that a linearized stress tensor damps Alfv\'en waves only negligibly, while damping magnetoacoustic waves significantly, was further reinforced by related work by \citet{1996Oughton,1997Oughton}.

Similar ideas to ours regarding the importance of nonlinearity were advocated by \citet{1986Nocera}, who modelled Alfv\'en waves subject to the $\eta_0$ part of the Braginskii viscous stress tensor, retaining the leading-order nonlinear terms in the wave perturbations. Consistent with the argument above, their calculations found that coronal Alfv\'en waves damp nonlinearly by parallel viscosity. The current paper complements and extends the previous analysis by \citet{1986Nocera}, with the goal of producing a comprehensive understanding of the nonlinear damping and field-aligned flows of propagating shear Alfv\'en waves with Braginskii viscosity.

A limitation of the mathematical techniques used in this paper is that they exclude certain other nonlinear effects that may be important in plasmas, such as nonlinear interactions between waves. Numerical investigations will be required in future to verify the analytical theory presented here, compare the relative importance of viscous damping and other nonlinear effects such as parametric decay instability, and consider interactions between nonlinear processes in Braginskii MHD.

This paper is organised as follows. \S~\ref{sec:bg} provides scientific background on single-fluid Braginskii MHD and its relationship to other single-fluid plasma models. \S~\ref{sec:heating_analyses} quantitatively examines Alfv\'en wave heating by the full Braginskii viscous stress tensor, demonstrating the importance of nonlinear $\eta_0$ terms and compressibility, and obtaining the wave decay properties for the weakly viscous limit using energy principles.
In \S~\ref{sec:low_Re}, we argue that in highly viscous limit, viscous heating is suppressed by self-organisation of the ponderomotive flow, which implies that viscosity strongly alters the field-aligned flow associated with Alfv\'en waves in this regime.
\S~\ref{sec:damping} further strengthens the analysis, using multiple scale analysis to obtain the decay properties without restrictions on the Alfv\'enic Reynolds number, assuming the framework of Braginskii MHD. The paper finishes with discussion in \S~\ref{sec:disc} and summary of main conclusions in \S~\ref{sec:conc}.

\section{Braginskii MHD}\label{sec:bg}

Braginskii MHD is an important plasma description that treats anisotropic viscosity and thermal conduction using rigorous closure from physical principles. This section provides a short primer on single-fluid Braginksii MHD, its connection with pressure (or temperature) anisotropy, and its relation to classical MHD and the CGL double-adiabatic equations.

As is described in various plasma textbooks, fluid variables can be rigorously and robustly defined as velocity moments of the underlying particle distribution functions. Transport equations for each particle species are then derived by taking moments of the kinetic Boltzmann equation, and combined to obtain the single fluid equations. Recommended presentations can be found in \citet{2009SchunkNagy} Chapter 7 and the Appendix of \citet{1962Spitzer}. 

Assuming quasi-neutrality, and conservation of mass, momentum and energy, this process yields the mass continuity equation,
\begin{equation}\label{eq:mass}
    \frac{\partial \rho}{\partial t} + \nabla\cdot\left(\rho \mathbf{V}\right) = 0,
\end{equation}
momentum equation
\begin{equation}\label{eq:mom_pressure_tensor}
    \rho \frac{\mathrm{D}\mathbf{V}}{\mathrm{D}t}
    = -\nabla\cdot\mathbf{P}+\rho\mathbf{G}+\mathbf{j}\times\mathbf{B},
\end{equation}
energy equation,
\begin{equation}\label{eq:energy_full}
    \frac{\mathrm{D}}{\mathrm{D}t}\left(\frac{3}{2}p\right) + \frac{5}{2}p\nabla\cdot\mathbf{V} = - \mathbf{ \pi}:\nabla \mathbf{V}  -\nabla\cdot\mathbf{q}+\mathbf{j}\cdot\left(\mathbf{E}+\mathbf{V}\times\mathbf{B}\right),
\end{equation}
higher-order transport equations if required, and the generalized Ohm's law.

The pressure tensor $\mathbf{P}$ that appears in Eq.~(\ref{eq:mom_pressure_tensor}) is the most fundamental representation of the internal forces associated with thermal motions of particles. It is symmetric, so it represents six degree of freedom. The momentum equation can also be reformulated by introducing the scalar pressure and stress tensor as
\begin{align}\label{eq:p_and_pi}
    p &= \frac{1}{3}\operatorname{Trace}\left(\mathbf{P}\right) = \frac{1}{3}P_{\alpha\alpha}, &
    \pi_{\alpha\beta} &= P_{\alpha\beta}-p\delta_{\alpha\beta},
\end{align}
where $\delta_{\alpha\beta}$ is the Kronecker delta.
So defined, the stress tensor $\mathbf{\pi}$ is symmetric and traceless.
These definitions gives the replacement $-\nabla\cdot\mathbf{P}=-\nabla p -\nabla\cdot\mathbf{\pi}$.

Deriving transport equations by moment taking meets with a fundamental closure problem: the transport equation for each fluid variable depends on a higher-order variable, producing an infinite regress unless the system can be closed by other considerations. The method of closure is therefore a major distinguishing feature between different fluid models for plasmas. It is also a major source of validity caveats. Various different methods of closure produce governing equations that conserve mass, momentum and energy, since these properties are already built into Eqs.~(\ref{eq:mass})--(\ref{eq:energy_full}). However, the different models discussed below disagree on the internal forces and heating, and can therefore produce different behaviors.

Classical MHD \citep{1937Hartmann,1942Alfven,1943Alfven,1950Batchelor} corresponds to a closure treatment in which the stress tensor and the heat flow vector are dropped from Eqs.~(\ref{eq:mom_pressure_tensor}) and (\ref{eq:energy_full}). Dropping the stress tensor can be justified when particle collisions or other forms of particle scattering such as wave-particle interactions are frequent enough that the pressure tensor remains very close to isotropic. The resulting MHD equations are valid for many situations, for instance modelling static equilibria, or dynamic situations in which the divergence of the stress tensor remains small compared to the Lorentz force. It is nonetheless a truncation since higher order variables are set to zero rather than approximated. Furthermore, collisionality in environments such as the solar corona is low enough that the stress tensor can become significant for various dynamic phenomena, including MHD waves.

Braginskii MHD uses a less restrictive method of closure. As is detailed by \citet{1965Braginskii}, when the collisional mean free path is significantly shorter than length scales over which fluid quantities vary, the heat flow vector takes the form of an anisotropic thermal conduction, and the stress tensor takes the form of an anisotropic viscosity. Closure can therefore be achieved by expressing $\mathbf{q}$ and $\pi$ in terms of lower-order fluid variables, which are traditionally derived using methods similar to \citet{1939ChapmanCowling} or \citet{1949Grad}.

The anisotropy inherent in $\mathbf{q}$ and $\pi$ can be appreciated heuristically, by considering the helical motion of charged particles in magnetized plasmas. The mean free path parallel to the magnetic field is the same as for unmagnetized plasmas, implying that transport parallel to the magnetic field is the same as for unmagnetized plasmas. Meanwhile, the mean free path perpendicular to the magnetic field is the gyroradius, which is typically much less than the mean free path parallel to the magnetic field, which supresses perpendicular transport. Hence both thermal conduction and viscous stresses are anisotropic with respect to the magnetic field direction, often extremely so.

The full Braginskii stress tensor, used in \S~\ref{sec:heating_analyses}, involves five viscosity coefficients. A useful simplification, used in \S~\ref{sec:damping}, is that for strong magnetizations, $\Omega_i \tau_i \gg 1$, the parallel $\eta_0$ coefficient greatly exceeds the other viscosity coefficients. Hence, one can often simplify by neglecting the smaller coefficients (although, as shown in \S~\ref{sec:heating_analyses} it can be necessary to retain other viscosity coefficients if length scales are highly anisotropic). In this simplification, one has the following covariant expressions for parallel viscosity \citep{1981Lifshitz,1986Hollweg}:
\begin{equation}
\begin{aligned}
    \pi_{\alpha\beta} &= -3\eta_0\left(h_\alpha h_\beta - \frac{\delta_{\alpha\beta}}{3}\right)
    \left(h_\mu h_\nu - \frac{\delta_{\mu\nu}}{3}\right)\partial_\mu V_\nu,
    \label{eq:pi_eta0}\\
    Q_{visc} &= 3 \eta_0\left(\left(h_\alpha h_\beta - \frac{\delta_{\alpha\beta}}{3}\right)\partial_\alpha V_\beta\right)^2,
\end{aligned}
\end{equation}
where $\mathbf{h}=\mathbf{B}/|\mathbf{B}|$ is the unit vector in the direction of the magnetic field.
These expressions are different to the isotropic viscosity that appears in the Navier-Stokes equations, owing to the anisotropy introduced by the magnetic field.

Parallel viscosity is closely related to pressure anisotropy. As pointed out by \citet{1956Chew}, when $\Omega_i \tau_i \gg 1$ the particle Lorentz force makes the pressure tensor gyrotropic, giving it the form
\begin{equation}\label{eq:gyro_pressure}
    P_{\alpha\beta} = p_\perp \delta_{\alpha\beta} + (p_{||}-p_\perp)h_\alpha h_\beta.
\end{equation}
This is a significant simplification, since the six degrees of freedom of a general pressure tensor have been replaced with two variables, $p_{||}$ and $p_\perp$.
The definitions in Eqs.~(\ref{eq:p_and_pi}) then yield $p=(p_{||}+2p_\perp)/3$ and 
\begin{align}\label{eq:pi_gyro}
 \pi_{\alpha\beta}&=(p_{||}-p_\perp)\left(h_\alpha h_\beta-\frac{\delta_{\alpha\beta}}{3} \right).
\end{align}

Equation~(\ref{eq:pi_gyro}) shows that pressure anisotropy has an equivalent stress tensor, which is proportional to $p_{||}-p_\perp$. Furthermore, Eqs.~(\ref{eq:pi_eta0}) and (\ref{eq:pi_gyro}) both have the form $\pi_{\alpha\beta}\sim(h_\alpha h_\beta - \delta_{\alpha\beta}/3)$, so equivalence of the stress tensors reduces to equivalence of the scalar factors in the two equations. An illuminating analysis of the conditions under which they converge has been written by \citet{1985Hollweg,1986Hollweg}, the most important condition being that collisions (or other processes such wave-particle interactions) relax the pressure anisotropy driven by velocity gradients to an extent that the pressure is only weakly anisotropic. Classical MHD, for comparison, assumes that pressure anisotropy can be neglected altogether.

For low collisionality, the quasistatic approximation in Braginskii MHD ceases to be valid and strong pressure anisotropy may develop. Under these conditions, separate evolution equations can be derived for $p_{||}$ and $p_{\perp}$ \citep{1956Chew,1986Hollweg}. However, the closure problem rears its head again, because those equations depend on the heat flow vector. A simple approach to obtaining a closed system is to ignore the heat flow vector, thus obtaining the CGL double adiabatic equations \citep{1956Chew}, which are commonly used for collisionless plasma. More sophisticated approaches also exist that solve for the evolution of the pressure anisotropy or the evolution of the stress tensor, retaining the heat flow vector and closing by other means.
The works by \citet{1988Balescu,2009SchunkNagy,2014Zank,2019HunanaPart1,2019HunanaPart2,2022Hunana} provide further reading on this topic.

Summarising, there exists a family of adjacent (sometimes overlapping) single-fluid models for plasmas. The most appropriate choice for a particular problem and/or context depends on the collisionality. When MHD timescales are greater than the ion collision time, Braginskii MHD provides rigorous closure and treats the internal forces and heat flow more accurately than classical MHD.

\section{Alfv\'en Wave Heating by Braginskii Viscosity}\label{sec:heating_analyses}

\subsection{Model}\label{sec:model}

We quantitatively examine the viscous dissipation for an Alfv\'en wave, which is a transverse wave polarized so that the magnetic perturbation is perpendicular to the equilibrium magnetic field and the wavevector. Setting the equilibrium magnetic field in the $z$-direction, the magnetic perturbation in the $x$-direction and the wavevector in the $yz$-plane, we consider a total magnetic field of the form
\begin{equation}\label{eq:aw:b}
    \mathbf{B} = b(y,z,t)\mathbf{e}_x + B_0\mathbf{e}_z.
\end{equation}
This ansatz automatically satisfies $\nabla\cdot\mathbf{B}=0$.
For the velocity field we assume the form
\begin{equation}\label{eq:aw:v}
    \mathbf{V} = V_x(y,z,t)\mathbf{e}_x + V_z(y,z,t)\mathbf{e}_z.
\end{equation}
The $V_x$ is the dominant velocity component. In linearized theory it would be the only component of $\mathbf{V}$. Additionally, we have explicitly included a higher-order $V_z$ term that represents the nonlinear ponderomotive flow parallel to the equilibrium magnetic field, which is driven by gradients of the magnetic pressure perturbation $b^2/2\mu_0$ associated with a finite-amplitude Alfv\'en wave \citep[e.g.][]{1971Hollweg}. The $V_z$ term can be dropped when the plasma is incompressible (see \S~\ref{sec:incomp}), however it is required for a nonlinear treatment of compressible plasma and affects the wave heating via the parallel viscosity coefficient $\eta_0$ (as remarked in \S~\ref{sec:intro}). The expression for $V_z$ in classical MHD is given later in Eq.~(\ref{eq:finite_beta:Vz}).

In a full solution, derivatives of $b^2/2\mu_0$ with respect to $y$ give rise to an additional nonlinear $y$-component of $\mathbf{V}$, which in turn produces a nonlinear $y$-component of $\mathbf{B}$. These terms are not shown explicitly in Eqs.~(\ref{eq:aw:b}) and (\ref{eq:aw:v}). Such terms were included by \citet{1986Nocera} and appear not to affect our main conclusions, provided the perpendicular wavelength of the Alfv\'en wave is sufficiently large.

The viscous force is determined from the viscous stress tensor $\pi_{\alpha\beta}$ by 
\begin{equation}\label{eq:F_general}
F_{visc,\alpha} = -\frac{\partial \pi_{\alpha\beta}}{\partial x_\beta},
\end{equation}
and the viscous heating rate is determined using
\begin{equation}\label{eq:Q_general}
    Q_{visc} = -\pi_{\alpha\beta} \frac{\partial V_\alpha}{\partial x_\beta},
\end{equation}
where $\alpha \in \left\{x, y, z\right\}$, $\beta \in \left\{x, y, z\right\}$, the $x_\beta$ are components of the position vector, $V_\alpha$ are components of $\mathbf{V}$ and repeated indices imply summation in the Einstein convention.

A vital point is that the viscous stress tensor depends on the direction of the magnetic field given by the unit vector $\mathbf{h} = \mathbf{B}/|\mathbf{B}|$, which for our Alfv\'en wave model in Eq.~(\ref{eq:aw:b}) has 
\begin{equation}\label{eq:h}
h_x = \frac{b}{\sqrt{B_0^2+b^2}}, \quad
h_y = 0, \quad
h_z = \frac{B_0}{\sqrt{B_0^2+b^2}},
\end{equation}
with $h_x^2 + h_z^2 = 1$. Our analysis differs from many past works by considering ${h_x\neq0}$ and identifying the dominant heating contribution at the end, as opposed to setting $h_x = 0$ before evaluating the damping effect on Alfv\'en waves.

Applying formulas from \S4 of \citet{1965Braginskii} \citep[equivalent matrix expressions are given by][]{1984Hogan},
the stress tensor is related to five viscosity coefficients by
\begin{equation}\label{eq:pi}
\pi_{\alpha\beta} = -\sum_{i=0}^2\eta_i W_{i\alpha\beta} 
  + \sum_{i=3}^4 \eta_i W_{i\alpha\beta}.
\end{equation}
The gyroviscous $\eta_3$ and $\eta_4$ terms do not contribute to heating, so evaluating the heating rate $Q_{visc}$ requires
\begin{equation}
\begin{aligned}
	W_{0\alpha\beta} &= \frac{3}{2}\left(h_\alpha h_\beta - \frac{1}{3}\delta_{\alpha\beta}\right)\left(h_\mu h_\nu - \frac{1}{3} \delta_{\mu\nu}\right)W_{\mu\nu}, \\
 	W_{1\alpha\beta} &= \left(\delta_{\alpha\mu}^\perp \delta_{\beta\nu}^\perp + \frac{1}{2} \delta_{\alpha\beta}^\perp h_\mu h_\nu \right)W_{\mu\nu}, \\
 	W_{2\alpha\beta} &= \left(\delta_{\alpha\mu}^\perp h_\beta h_\nu + \delta_{\beta\nu}^\perp h_\alpha h_\mu \right)W_{\mu\nu},
 \end{aligned}
\end{equation}
where $\delta_{\alpha\beta}$ is the Kronecker delta,
\begin{equation}
   \delta_{\alpha\beta}^\perp = \delta_{\alpha\beta}-h_\alpha h_\beta, 
\end{equation}
and the rate of strain tensor is
\begin{equation}
    W_{\alpha\beta}= \frac{\partial V_\alpha}{\partial x_\beta}
    +\frac{\partial V_\beta}{\partial x_\alpha}
    -\frac{2}{3} \delta_{\alpha\beta}\nabla\cdot\mathbf{V}.
\end{equation}

For the shear Alfv\'en wave geometry described by Eq.~(\ref{eq:aw:v}), the $\mathbf{W}_{i}$ tensors become
\begin{eqnarray}
    \mathbf{W}_0 &=& 
    \left(h_x h_z \partial_z V_x
    +\left(\tfrac{2}{3}-h_x^2\right) \partial_z V_z\right)
    \begin{pmatrix}
        3h_x^2-1 &  0 & 3h_x h_z \\
            0    & -1 &     0    \\
        3h_x h_z  &  0 & 3h_z^2-1
    \end{pmatrix}, 
    \label{eq:W0} \\
    \mathbf{W}_1 &=& 
    h_x \left(h_z \partial_z V_x-h_x \partial_z V_z \right)
    \begin{pmatrix}
        -h_z^2  & 0 & h_x h_z \\
           0    & 1 &   0     \\
        h_x h_z & 0 & -h_x^2
    \end{pmatrix} 
    +\left(h_x \partial_y V_x-h_z \partial_y V_z\right)
    \begin{pmatrix}
         0  & h_z  &   0  \\
        h_z &   0  & -h_x \\
         0  & -h_x &   0
    \end{pmatrix},
    \\
    \mathbf{W}_2 &=& 
    \left(\left(1-2h_x^2\right)\partial_z V_x - 2h_xh_z \partial_z V_z\right)
    \begin{pmatrix}
        2h_x h_z & 0 & 1-2h_x^2 \\
             0   & 0 &     0    \\
        1-2h_x^2 & 0 & -2h_x h_z
    \end{pmatrix} 
    +\left(h_x \partial_y V_x + h_z \partial_y V_z\right)
    \begin{pmatrix}
         0  & h_x &  0 \\
        h_x &  0  & h_z \\
         0  & h_z &  0
    \end{pmatrix}, \label{eq:W2}
\end{eqnarray}
The viscous heating rate with $h_x\neq0$ retained is thus
\begin{multline}\label{eq:Q}
    Q_{visc}=
    \frac{\eta_0}{3} \left(3h_x h_z \partial_z V_x+(2-3h_x^2) \partial_z V_z \right)^2 
    +\eta_1 h_x^2 \left( h_z \partial_z V_x-h_x \partial_z V_z \right)^2
    +\eta_1 \left( h_z \partial_y V_x - h_x \partial_y V_z\right)^2 
    \\
    +\eta_2 \left(\left(1-2h_x^2\right)\partial_z V_x - 2h_x h_z \partial_z V_z \right)^2 
    +\eta_2 \left( h_x \partial_y V_x + h_z \partial_y V_z \right)^2.
\end{multline}

\subsection{Two small parameters}\label{sec:params}

As anticipated in \S~\ref{sec:intro} (e.g. Fig.~\ref{fig:reasoning_paths}), two parameters determine the relative importance of individual terms in Eq.~(\ref{eq:Q}). The first small parameter is $\left(b/B_0\right)^2$, the ratio of the wave’s magnetic energy density to the energy density of the background magnetic field, which enters through $h_x$ and $h_z$. In the modern era, extensive observations of coronal MHD waves \citep{2005Nakariakov,2012DeMoortel} allow $\left(b/B_0\right)^2$ to be quantified with good certainty, directly from resolved wave observations or indirectly from spectral line widths. For example, \citet{2015Morton} studied waves at the base of a coronal open field region using both approaches and reported a wave speed $v_A = 400\mbox{ km s}^{-1}$ and wave motions at $v = 35\mbox{ km s}^{-1}$. Both measurements are consistent with earlier findings for coronal holes and the quiet Sun \citep[e.g.][]{2011McIntosh}. 
From observations like these, $h_x^2 \sim \left(b/B_0\right)^2 \sim \left(v/v_A\right)^2 \sim 10^{-2}$ .

The second small parameter is $\left(\Omega_i \tau_i\right)^{-2}$, which sets the viscosity coefficients $\eta_1$ and $\eta_2$ relative to $\eta_0$. 
The value of $\Omega_i \tau_i$ can vary significantly in the corona, but if magnetic null points are excluded one obtains values similar to the estimates made by \citet{1985Hollweg}, who found $3.4\times10^5$ for a solar active region and $7.2\times10^5$ near the base of a coronal hole. We therefore expect $(\Omega_i \tau_i)^{-2} \lesssim 10^{-11}$ under common conditions, and $\eta_1$ and $\eta_2$ simplify to
\begin{equation}\label{eq:eta_perp}
 	\eta_2 = \frac{6408}{5125}\left(\Omega_i \tau_i\right)^{-2}\eta_0, \quad 
 	\eta_1 = \frac{1}{4}\eta_2.
\end{equation}
The numerical coefficients in Eq.~(\ref{eq:eta_perp}) are obtained in the limit $(\Omega_i \tau_i)^{-2}\to 0$, e.g. from Eq.~(73) of \cite{2022Hunana}. They are approximate for finite $(\Omega_i \tau_i)^{-2}$ but have a high degree of accuracy because the corrections to the coefficients are of the order of $(\Omega_i \tau_i)^{-2}\lesssim 10^{-11}$. Inspecting Eq.~(\ref{eq:eta_perp}) and considering $(\Omega_i \tau_i)^{-2}\lesssim 10^{-11}$, the $\eta_2$ and $\eta_1$ coefficients are both vastly smaller than $\eta_0$.

The smallness of $\left(b/B_0\right)^2$ and the smallness of $\left(\Omega_i \tau_i\right)^{-2}$ compete to make different terms dominate the viscous heating. If one tries to simplify Eq.~(\ref{eq:Q}) by setting $\left(b/B_0\right)^2$ to zero, then $h_x = 0$ and $V_z = 0$ gives 
$Q_{visc}= \eta_2 \left(\partial_z V_x\right)^2 + \eta_1 \left(\partial_y V_x\right)^2$, as obtained by \citet{1965Braginskii}. On the other hand, if one tries to simplify by first taking $\left(\Omega_i \tau_i\right)^{-2}$ to zero then only $\eta_0$ terms remain, suggesting a different conclusion. Thus, the quantitative results recover the two branches shown in Fig.~\ref{fig:reasoning_paths}. 
To correctly determine the damping under coronal conditions, one must carefully compare terms in the full Eq.~(\ref{eq:Q}), bearing in mind that there are two small parameters, which we do now (diagonal branch in Fig.~\ref{fig:reasoning_paths}).

\subsection{Heating rate for incompressible plasma}\label{sec:incomp}

The analysis for incompressible plasma is relatively straightforward, which makes it a natural starting point for discussion. The assumption of incompressibility is appropriate for liquid metals or high-beta plasmas, but not, we note, for the corona. The use of coronal wave amplitudes and magnetizations in this section is therefore intended to be instructive only, with the compressible finite-beta treatment that follows later in this paper being required to treat the corona.

In the incompressible case, $\nabla\cdot\mathbf{V}=0$ applied to our Alfv\'en wave geometry implies $V_z = 0$. Thus Eq.~(\ref{eq:Q}) with $\eta_1 = \frac{1}{4}\eta_2$ simplifies to 
\begin{equation}
\label{eq:incomp:Q}
Q_{visc}=\left\{3\eta_0 h_x^2 h_z^2+\eta_2 \left(\left(1-2h_x^2 \right)^2+\frac{1}{4} h_x^2 h_z^2 \right)\right\} \left(\partial_z V_x \right)^2 \\
+\eta_2 \left(\frac{1}
{4}h_z^2+h_x^2 \right) \left(\partial_y V_x\right)^2.
\end{equation}

The terms involving $\partial_z V_x$ set the viscous dissipation due to wavelengths parallel to the equilibrium magnetic field, and we first ask whether dissipation due to parallel wavelengths is dominated by the linear $\eta_2$ contribution that has been widely recognised since \citet{1965Braginskii}, or the nonlinear $\eta_0$ contribution. The ratio of the two terms inside the curly brackets in Eq.~(\ref{eq:incomp:Q}) is
\begin{equation}\label{eq:incomp:par_ratio}
3h_x^2 \frac{\eta_0}{\eta_2}\left[\frac{1-h_x^2}{(1-2h_x^2)^2+\frac{1}{4}h_x^2 (1-h_x^2)}\right]
\approx 3h_x^2 \frac{\eta_0}{\eta_2}
\approx 2.4 h_x^2 (\Omega_i \tau_i )^2, 
\end{equation}
where the first step simplifies using $h_x^2\ll1$ (for the observed value of $h_x^2\approx10^{-2}$, retaining the terms in the square bracket increases the ratio by 2.8\%, so this approximation is both accurate and conservative) and the substitution for $\eta_0/\eta_2$ is by Eq.~(\ref{eq:eta_perp}). For the coronal wave amplitudes and $\Omega_i\tau_i$ values noted in \S~\ref{sec:params}, this ratio exceeds $10^9$, with the nonlinear damping via $\eta_0$ dominating the heating rate by that factor. In other words, the viscous dissipation of Alfv\'en waves via derivatives aligned with the equilibrium magnetic field is a factor $10^9$ stronger than predicted by linear theory.

We now evaluate the role of derivatives perpendicular to the equilibrium magnetic field by comparing the nonlinear $\eta_0$ term in Eq.~(\ref{eq:incomp:Q}) to the term involving $\partial_y V_x$. The ratio of these heating rate terms is
\begin{equation}\label{eq:incomp:perp_ratio}
12h_x^2  \frac{\eta_0}{\eta_2}  \left(\frac{\lambda_\perp}{\lambda_{||}}\right)^2 \left[\frac{1-h_x^2}{1+3h_x^2}\right]
\approx 12h_x^2  \frac{\eta_0}{\eta_2} \left( \frac{\lambda_\perp}{\lambda_{||}} \right)^2 
\approx 9.6h_x^2 \left( \Omega_i \tau_i \right)^2 \left( \frac{\lambda_\perp}{\lambda_{||}} \right)^2,
\end{equation}
where $\lambda_\perp$ and $\lambda_{||}$ are the wavelengths perpendicular and parallel to the equilibrium magnetic field (for the observed value of $h_x^2\approx10^{-2}$, the  approximation of the terms in $h_x^2$ is accurate to 3.9\%). For the coronal parameters noted above, if $\lambda_\perp \approx\lambda_{||}$ then the nonlinear $\eta_0$ term again dominates by a factor that exceeds $10^9$. For smaller transverse wavelengths, the nonlinear $\eta_0$ term dominates whenever $\lambda_\perp \gtrsim 10^{-5}\lambda_{||}$. If one considers a wave speed of $400\mbox{ km s}^{-1}$ and a frequency of 3~mHz, consistent with the observations by \citet{2015Morton}, the condition that the nonlinear $\eta_0$ term dominates becomes $\lambda_\perp \gtrsim 800\mbox{ m}$. Given that CoMP has imaged Alfv\'enic waves using 3~Mm pixels, this condition appears to be met by a very large margin, making the nonlinear $\eta_0$ dissipation dominant over the $\eta_2$ linear dissipation.

The purpose of deriving Eq.~(\ref{eq:incomp:Q}) and the ratios on the left hand sides of Eq.~(\ref{eq:incomp:par_ratio}) and (\ref{eq:incomp:perp_ratio}) such that they include all appearances of $h_x$ and $h_z$ is that they can be evaluated exactly for a given value of $h_x$. This makes it explicit that our conclusions are insensitive to the precise value of $h_x$, only that the value of $h_x$ is broadly consistent with coronal observations. While that approach is most comprehensive, the same conclusions can also be reached by separately simplifying each term in Eq.~(\ref{eq:incomp:Q}) using $h_x^2\ll 1$ and $h_z^2=1-h_x^2\approx 1$ to obtain the less cumbersome formula 
\begin{equation}\label{eq:incomp:Q_simple}
    Q_{visc}=\left\{3\eta_0 h_x^2 +\eta_2 \right\} \left(\partial_z V_x \right)^2+\eta_1 \left(\partial_y V_x\right)^2,
\end{equation}
and comparing terms to reach the same conclusions.

\subsection{Compressible plasma with large Re}\label{sec:compressible}
Under typical coronal conditions, the thermal pressure is too small to prevent compression of the plasma by nonlinear magnetic pressure forces, thus a nonlinear $V_z$ develops that is known as the ponderomotive flow \citep{1971Hollweg}. This flow component affects the viscous heating rate via the parallel viscosity coefficient $\eta_0$, hence compressible theory is required for nonlinear viscous damping of Alfv\'en waves in plasma.

We define the Alfv\'enic Reynolds number as
\begin{equation}\label{eq:Re}
    \mathrm{Re}=\frac{\rho v_A}{k_{||} \eta_0}.
\end{equation}
This dimensionless parameter differs from the traditional Reynolds number since it refers to the Alfv\'en speed $v_A=B/\sqrt{\mu_0 \rho}$ instead of a typical fluid velocity. This distinction mirrors that between the Lundquist number and magnetic Reynolds number in resistive MHD. Justification for defining $\mathrm{Re}$ according to Eq.~(\ref{eq:Re}) will be found in the detailed mathematical solutions in \S~\ref{sec:damping}, in which it is found to be a natural parameter of the system \citep[also see][]{1986Nocera}.

In this section, the ponderomotive $V_z$ will be related to $V_x$ using expansions in the amplitude of the primary wave fields. Several assumptions are used to accomplish this. First, we make use of the result that a travelling wave solution propagating in the positive $z$-direction has 
\begin{equation}\label{eq:travelling_derivs}
    \frac{\partial}{\partial t} \equiv -v_A\frac{\partial}{\partial z},
\end{equation}
where $v_A$ is the wave speed.
For simplicity it is assumed that derivatives of background quantities are sufficiently weak to play a higher-order role on the dynamics.
We also simplify here by replacing full treatment of thermal conduction with two thermodynamic cases: adiabatic and isothermal.
Finally, it is assumed that $\mathrm{Re}$ is large enough that viscous forces can be neglected at leading order when evaluating $V_z$, which makes it possible to obtain an algebraic relationship between $V_z$ and $b^2$. This assumption will be removed for \S~\ref{sec:damping}, in which the effect of viscous forces on $V_x$ and $V_z$ is included. 

The $x$-components of the momentum and induction equations are unaffected by the ponderomotive flow at leading order in the wave amplitude. From them one recovers the Wal\'en relation for propagating Alfv\'en waves, $b/B_0 = -V_x/v_A$.

At leading order, the $z$-component of the momentum equation is
\begin{equation}\label{eq:mom_z}
    \rho_0\frac{\partial V_z}{\partial t}+\frac{\partial}{\partial z}\left(\delta p+\frac{b^2}{2\mu_0}\right)=0,
\end{equation}
where the viscous force has been neglected since we currently consider the limit of large $\mathrm{Re}$.
Using Eq~(\ref{eq:travelling_derivs}) and integrating yields an algebraic relationship between $V_z$, $\delta p$ and $b^2$.
In an adiabatic treatment, the energy equation yields $\delta p = \gamma p_0 V_z/v_A$, hence we obtain
\begin{equation}\label{eq:finite_beta:Vz}
    \frac{V_z}{v_A} = \frac{1}{2(1-\beta)}\left(\frac{b}{B_0}\right)^2,
\end{equation}
where
\begin{equation}
    \beta = \left(\frac{c_s}{v_A}\right)^2 = \frac{\gamma}{2}\left(\frac{p}{B_0^2/2\mu_0}\right).\label{eq:beta}
\end{equation}

In an isothermal treatment, the ideal gas law $p=\rho R T$ yields $\delta p / p_0 = \delta \rho / \rho_0 = V_z/v_A$. This does not change the form of Eq.~(\ref{eq:finite_beta:Vz}); instead, the isothermal case is recovered simply by setting $\gamma=1$ in the definition of $\beta$.

Defining $\beta$ as the square of the ratio of the sound speed ($c_s = \sqrt{\gamma p_0/\rho_0}$) to the Alfv\'en speed differs slightly from the convention of defining $\beta$ as the ratio of thermal pressure to magnetic pressure, due to the factor $\gamma/2$, which is 5/6 for an adiabatic monoatomic gas, and $1/2$ for an isothermal model. Defining $\beta$ as the speed ratio squared leads to cleaner mathematics for many MHD wave problems, including this one, and it has therefore become established practice in MHD wave theory. 

The $\beta=1$ singularity in Eq.~(\ref{eq:finite_beta:Vz}) arises because  $c_s=v_A$ implies resonance between the Alfv\'en wave and an acoustic wave, which resonantly transfers energy between the waves. In this specific case, Eq.~(\ref{eq:travelling_derivs}) does not apply because it does not account for evolution due to resonance. 
Similarly, Eq.~(\ref{eq:mom_z}) assumes that $V_z\ll v_A$ to simplify the convective derivative, and the solution in Eq.~(\ref{eq:finite_beta:Vz}) does not satisfy this condition in the immediate vicinity of $\beta=1$.
The $\beta=1$ resonance and the singularity in Eq.~(\ref{eq:finite_beta:Vz}) are not of concern for most coronal applications, which typically have $\beta<0.2$, but there are special cases in which it is of interest, such as waves propagating towards coronal magnetic nulls or across the $\beta=1$ layer in the lower solar atmosphere. \citet{2016Russell} have previously applied such nonlinear resonant coupling to the problem of sunquake generation by magnetic field changes during solar flares.

Differentiating Eq.~(\ref{eq:finite_beta:Vz}) and employing the relation $b/B_0 = -V_x/v_A$ yields
\begin{equation}\label{eq:finite_beta:Vz_deriv}
    \partial_\alpha V_z 
    = -\frac{1}{(1-\beta)}\frac{h_x}{h_z} \partial_\alpha V_x,
\end{equation}
which can be used to eliminate $V_z$ from Eq.~(\ref{eq:Q}).
To leading order in $h_x^2$ in each viscosity coefficient, we find
\begin{equation}\label{eq:finite_beta:Q}
 Q_{visc}= \left\{C\eta_0h_x^2
 +\eta_2 \right\}\left(\partial_z V_x \right)^2
 +\eta_1 (\partial_y V_x)^2,
\end{equation}
where
\begin{equation}\label{eq:C}
    C=\frac{1}{3} \left(\frac{1-3\beta}{1-\beta}\right)^2.
\end{equation}

Some special cases are noteworthy. The incompressible results of \S~\ref{sec:incomp} are recovered for $\beta\to\infty$, which gives $V_z\to0$ and $C\to 3$. Similarly, the cold plasma solution is recovered by setting $\beta=0$, which gives  $V_z=\left(b/B_0\right)^2/2$ and $C=1/3$. 

The nonlinear heating rate for cold plasma ($\beta=0$) is a factor nine smaller than for incompressible plasma ($\beta\to\infty$), which demonstrates the importance of compressibility for this problem. Furthermore, $C(\beta)$ is monotonically decreasing between $\beta=0$ and $\beta=1/3$. Since $0<\beta<1/3$ for most coronal applications, the dissipation rate due to nonlinear Braginskii viscosity in these environments is reduced compared to the cold plasma solution. For example, given $\beta=0.1$, the heating rate is approximately 60\% of the value for cold plasma. It is therefore evident that compressibility and finite-beta effects must be treated when assessing viscous dissipation of Alfv\'en waves.

Another important feature is that $C$ has a zero for $\beta=1/3$. 
This is one circumstance in which
\begin{equation}\label{eq:comp:Vz_low_Q}
    \frac{V_z}{v_A}=\frac{3}{4}\left(\frac{V_x}{v_A}\right)^2 =\frac{3}{4}\left(\frac{b}{B_0}\right)^2,
\end{equation}
which causes cancellation within the $\eta_0$ contribution to $Q_{visc}$. That a particular organisation of $V_z/v_A$ can suppress nonlinear viscous dissipation is an important novel finding that \S~\ref{sec:low_Re} explores further in the context of low $\mathrm{Re}$.

The final feature of $C$ is the singularity at $\beta=1$. As noted earlier in this section, $c_s=v_A$ implies that the Alfv\'en wave is in resonance with a sound wave, which transfers energy between the Alfv\'en wave and the sound wave. Caution is needed around the resonance, since resonant energy transfer cannot be described using Eq.~(\ref{eq:travelling_derivs}), which was used to derive Eq.~(\ref{eq:C}).

Evaluation of ratios of heating terms from Eq.~(\ref{eq:finite_beta:Q}) proceeds as for the comparison in Sec.~\ref{sec:incomp}, but with $\eta_0$ multiplied by $C/3$. The top-level conclusions remain intact: heating by the Braginskii viscous stress tensor is dominated by an $\eta_0$ term that is nonlinear in the wave amplitude, and for coronal values, heating due to the nonlinear $\eta_0$ term is many orders of magnitude larger than the heating due to the linear $\eta_1$ and $\eta_2$ terms. 

\subsection{What wave amplitude is linear?}
An important implication of the preceding analysis is that nonlinear effects become significant for anisotropic viscosity at far lower wave amplitudes than they do for other terms in the MHD equations. Linearizing the Braginskii viscous stress tensor is only appropriate when $h_x^2 \sim (b/B_0)^2 \ll \left(\Omega_i \tau_i\right)^{-2}$, which in the corona corresponds to a requirement that the wave energy density is less than $10^{-11}$ times the energy density of the background magnetic field, far too small to be relevant to coronal energetics. Waves that have small enough amplitudes to be governed by linear viscous damping theory would be unobservable and have no effect on the coronal energy balance. Thus, for coronal Alfv\'en waves, viscosity must be treated nonlinearly in the wave amplitude, as well as anisotropically due to the magnetic field. Interestingly, this linearization condition is far more stringent than the linearization condition for other terms in the MHD equations, whereby $h_x^2$ is normally compared to unity. The extreme difference in these linearization conditions is due to the large $\eta_0/\eta_2$ ratio produced by the strong magnetization.

\subsection{Damping scales for large Re (energy derivation)}\label{sec:damping:simple}
It is of major interest to know the time and length scales over which waves damp. This section provides a relatively simple derivation of the decay scales for nonlinear viscous damping of propagating shear Alfv\'en waves for large $\mathrm{Re}$, using energy principles.

Dropping the $\eta_1$ and $\eta_2$ terms from Eq.~(\ref{eq:finite_beta:Q}), the heating rate due to the $\eta_0$ parallel viscosity coefficient for large $\mathrm{Re}$ is
\begin{equation}\label{eq:Q_eta0}
    Q_{visc}= \frac{\eta_0}{3}\left(\frac{1-3\beta}{1-\beta}\right)^2 \left(\frac{b}{B_0}\right)^2(\partial_z V_x )^2.
\end{equation}

A wave energy decay time can be defined according to
   $ \tau_d = \left<E_w\right>/\left<Q_{visc}\right>$,
where $\left<.\right>$ denotes the time average over a wave period, and $E_w$ is the wave energy density.
The corresponding decay length is $L_d=v_A \tau_d$.

For forward propagating Alfv\'en waves, $E_w \approx \rho_0 V_x^2$, and $V_x = a\cos(\phi)$ where $\phi=k_{||}(z-v_At)$. Hence,
\begin{equation}
    E_w = \rho_0 a^2 \frac{(1+\cos(2\phi))}{2}.
\end{equation}
Similarly, using Eq.~(\ref{eq:Q_eta0})
with $(b/B_0)^2\approx(V_x/V_A)^2$,
\begin{equation}
    Q_{visc}= \frac{\eta_0k_{||}^2}{3v_A^2}\left(\frac{1-3\beta}{1-\beta}\right)^2 a^4 \frac{(1-\cos(4\phi))}{8},
\end{equation}
which give the fast time averages
\begin{eqnarray}
    \left<E_w\right> &=& \frac{\rho_0 a^2}{2}, \\
    \left<Q\right> &=& \frac{\eta_0k_{||}^2a^4}{24v_A^2}\left(\frac{1-3\beta}{1-\beta}\right)^2.
\end{eqnarray}
The wave energy decay scales are therefore
\begin{eqnarray}
    \tau_d = \frac{12 \rho_0}{\eta_0 k_{||}^2 (a/v_A)^2}\left(\frac{1-\beta}{1-3\beta}\right)^2,
    \label{eq:damping:time} \\
    L_d = \frac{12 \rho_0 v_A}{\eta_0 k_{||}^2 (a/v_A)^2}\left(\frac{1-\beta}{1-3\beta}\right)^2.
    \label{eq:damping:length}
\end{eqnarray}

Equations~(\ref{eq:damping:time}) and (\ref{eq:damping:length}) show that waves with larger $k_{||}$ (equivalently, higher frequencies) are damped on shorter scales. We also remark that since $L_d$ depends on the amplitude of $V_x$ (the constant $a$), the decay envelope is non-exponential. 
The damping properties are elaborated on more fully in \S~\ref{sec:damping}, in which the assumption of large $\mathrm{Re}$ is removed and the functional form of the wave envelope is determined.

\section{Self-organised viscous flow}\label{sec:low_Re}
In the limit $\mathrm{Re}\to0$, the viscous force in the $z$-component of the momentum equation risk becoming extremely large, unless the flow self-organises to prevent this. Correspondingly, in the limit $\mathrm{Re}\to0$, strong dissipation will prevent waves from propagating, unless $V_z$ is determined by viscosity. One can therefore expect self-organisation of the flow pattern for Alfv\'en waves in highly viscous plasma (small values of $\mathrm{Re}$), which is a concept previously advanced by \citet{1992Montgomery}.

To investigate quantitatively, we analyze the highly magnetised regime $\Omega_i\tau_i\gg 1$, simplifying the stress tensor and heating rate by retaining only the $\eta_0$ parallel viscosity coefficient. Inspecting Eq.~(\ref{eq:pi_eta0}), components of $\pi_{\alpha\beta}$ are proportional to $(h_\mu h_\nu - \delta_{\mu\nu}/3)\partial_\mu V_\nu$, and $Q_{visc}$ is proportional to the square of this expression. Applying the shear Alfv\'en wave geometry of Eqs.~(\ref{eq:aw:b}) and (\ref{eq:aw:v}) and simplifying by $h_x^2\ll 1$,
\begin{equation}
    \left(h_\mu h_\nu - \frac{\delta_{\mu\nu}}{3}\right)\frac{\partial V_\nu}{\partial x_\mu}
    \approx h_x\frac{\partial V_x}{\partial z}+\frac{2}{3}\frac{\partial V_z}{\partial z}.
\end{equation}
Viscous forces and heating can be suppressed, allowing Alfv\'en wave propagation, if the flow self-organises to keep this expression close to zero. 
Using the Alfv\'en wave relation to substitute $h_x\approx b/B_0 \approx -V_x / v_A$ and integrating, we find that for small $\mathrm{Re}$
\begin{equation}\label{eq:Re_to_0:Vz_low_Q}
    \frac{V_z}{v_A}=\frac{3}{4}\left(\frac{V_x}{v_A}\right)^2=\frac{3}{4}\left(\frac{b}{B_0}\right)^2.
\end{equation}

The relation specified by Eq.~(\ref{eq:Re_to_0:Vz_low_Q}) appeared previously in the different context of \S~\ref{sec:compressible}, where it was seen that viscous dissipation of Alfv\'en waves in high $\mathrm{Re}$ plasma is suppressed for the special case of $\beta=1/3$. The flow pattern required to produce cancellation within the $\eta_0$ part of $Q_{visc}$ is independent of $\mathrm{Re}$ and $\beta$, but it occurs for different reasons in the two cases: in \S~\ref{sec:compressible} it arose as a special case of ponderomotive flow with finite $\beta$; when $\mathrm{Re}$ is small, it occurs because of self-organisation through viscous forces.

This novel result demonstrates that decay scales and other properties derived in \S~\ref{sec:heating_analyses} should not be extrapolated to small $\mathrm{Re}$. Instead, we expect that as $\mathrm{Re}\to0$, the viscous force organises the flow such that $V_z$ obeys Eq.~(\ref{eq:Re_to_0:Vz_low_Q}), for which  dissipation is suppressed by cancellation within the $\eta_0$ part of $Q_{visc}$.

\section{Multiple scale analysis}\label{sec:damping}

Section~\ref{sec:heating_analyses} used methods of analysis based on heating rates and energy principles. Section~\ref{sec:damping} now takes a complementary approach of solving the full set of governing equations using multiple scale analysis, to reinforce the results of \S~\ref{sec:heating_analyses}, extend to general $\mathrm{Re}$ by including the effect of the viscous force on $\mathbf{V}$, and obtain additional results including the functional form of the nonlinear decay.

\subsection{Comparison to Nocera et al. (1986)}\label{sec:nocera}

We preface the multiple scale analysis part of this paper with some remarks about related calculations by \citet{1986Nocera}. Their work and ours both concentrate on $\eta_0$ viscosity as the main source of wave damping, treating this nonlinearly in the wave amplitude (the horizontal branch of Figure~\ref{fig:reasoning_paths}). Also in common, both treat ponderomotive and finite $\beta$ effects.

The previous work of \citet{1986Nocera} derived a version of the viscous stress tensor that includes the leading order effect of $h_x\neq0$ in the $\eta_0$ term. Terms in the viscosity tensor were then compared, concluding like our \S~\ref{sec:heating_analyses} (but by different arguments) that the nonlinear $\eta_0$ term exceeds contributions from other viscosity coefficients when $(b/B_0)^2\gg(\Omega_i\tau_i)^{-2}$ (their Eq.~(3.13)). The two studies thus agree on the dominance of nonlinear $\eta_0$ viscosity.

\citet{1986Nocera} then found a decay length using the following strategy. A self-consistent perturbation ordering was introduced, then the $x$-components of the momentum and induction equations were combined to obtain a single equation for $V_x$, which at linear order is a wave equation. Next, all variables apart from $V_x$ were eliminated from the leading-order nonlinear term. Finally, they concluded from a stability analysis that waves with $k_\perp=0$ are damped nonlinearly, with a decay time that has the same form as our Eq.~(\ref{eq:damping:time}) (their Eq.~(5.7), given in terms of normalised variables).

The detailed derivation that follows in \S~\ref{sec:damping:detailed} draws inspiration from the framework developed by \citet{1986Nocera}. We have also taken the opportunity to make several changes that we regard as improvements, most importantly:
\begin{enumerate}

\item \citet{1986Nocera} assumed that the fast time average of $V_z$ is zero, which necessitated adding a non-zero constant of integration to $V_z$. By contrast, we will set the constant of integration to zero, which is the only choice for which an Alfv\'en wave driver switching on at one boundary does not unphysically send an instantaneous signal to infinity. Additional support for our choice comes from simulations of nonlinear longitudinal flows produced by Alfv\'en waves \citep[e.g.][]{2011McLaughlin}, which are consistent with the constraint used in our work.

\item The stability analysis in \S5 \citet{1986Nocera} is replaced with a multiple scale analysis of the type covered in Chapter~11 of \citet{1978BenderOrszag}.

\item \citet{1986Nocera} made their wave envelope a function of $z+v_A t$. We treat the envelope as time-independent and thus explicitly investigate damping of a propagating wave with respect to distance.

\item Our derivation provides the envelope of $V_x$ as well as the decay length.

\item Our solution is valid for general $\mathrm{Re}$, whereas \citet{1986Nocera} solved for the decay scales in the low-viscosity limit of high $\mathrm{Re}$ only.

\end{enumerate}

Equally, \citet{1986Nocera} treated cases that we do not, including the possibility of $k_\perp$ large enough for coupling between the Alfv\'en and fast modes to alter the wave properties (referred to in their paper as the case of phase mixed waves).

\subsection{Detailed solution}\label{sec:damping:detailed}

\subsubsection{Geometry and perturbations}
We assume the Alfv\'en wave geometry of Eqs.~(\ref{eq:aw:b}) and (\ref{eq:aw:v}), set $\partial/\partial y \equiv 0$ to concentrate on waves without short perpendicular scales, and introduce density and pressure perturbations $\delta\rho$ and $\delta p$ together with a self-consistent perturbation ordering that has $V_x/v_A\sim b/B_0\sim \epsilon^{1/2}$ and $V_z/v_A\sim \delta \rho/\rho_0\sim \delta p / p_0 \sim \epsilon$. The viscosity $\eta_0$ and background quantities $B_0$, $\rho_0$ and $p_0$ are treated as locally homogeneous for simplicity.

\subsubsection{Nonlinear wave equation}

Starting from the ideal induction equation, 
\begin{equation}
    \frac{\partial \mathbf{B}}{\partial t} = \nabla\times\left(\mathbf{V}\times\mathbf{B}\right),
\end{equation}
we have
\begin{equation}\label{eq:dbdt}
    \frac{\partial b}{\partial t} - B_0\frac{\partial V_x}{\partial z} =  - \frac{\partial}{\partial z}\left(b V_z\right)
    \quad \text{(exact)},
\end{equation}
where the linear terms have been grouped on the left hand side and the nonlinear term on the right hand side.

The momentum equation is
\begin{equation}
\label{eq:mom}
    \rho\left(\frac{\partial V_\alpha}{\partial t}+(\mathbf{V}\cdot\nabla) V_\alpha\right) = 
    -\frac{\partial}{\partial x_\alpha}\left(p+\frac{B^2}{2\mu_0}\right) 
    +\frac{1}{\mu_0}(\mathbf{B}\cdot\nabla)B_\alpha - \frac{\partial \pi_{\alpha\beta}}{\partial x_\beta}.
\end{equation}
When $\eta_0$ contributions dominate the viscous force, Eqs.~(\ref{eq:pi}) and (\ref{eq:W0}) give
\begin{equation}
    -\pi_{xz}=3\eta_0\frac{b}{B_0}\left(\frac{b}{B_0}\frac{\partial V_x}{\partial z}+\frac{2}{3}\frac{\partial V_z}{\partial z}\right) + O(\epsilon^{5/2})
\end{equation}
so the $x$-component of Eq.~(\ref{eq:mom}) becomes
\begin{equation}\label{eq:dVxdt}
    \frac{\partial V_x}{\partial t}-\frac{B_0}{\mu_0\rho_0}\frac{\partial b}{\partial z}
    = -\frac{\delta\rho}{\rho_0} \frac{\partial V_x}{\partial t}-V_z\frac{\partial V_x}{\partial z} 
    + \frac{3\eta_0}{\rho_0}\frac{\partial}{\partial z}\left(\frac{b}{B_0}\left[\frac{b}{B_0}\frac{\partial V_x}{\partial z}+\frac{2}{3}\frac{\partial V_z}{\partial z}\right]\right) + O(\epsilon^{5/2}),
\end{equation}
where linear terms and nonlinear terms have again been placed on opposite sides of the equation.

Taking the time derivative of Eq.~(\ref{eq:dVxdt}) and using Eq.~(\ref{eq:dbdt}) to eliminate $b$ from the linear terms,
\begin{equation}\label{eq:nonlinear_wave_eq_raw}
    \left(\frac{\partial^2}{\partial t^2}-v_A^2\frac{\partial^2}{\partial z^2}\right)V_x
    = -v_A^2 \frac{\partial^2}{\partial z^2}\left(\frac{b}{B_0} V_z\right) 
    -\frac{\partial}{\partial t}\left(\frac{\delta\rho}{\rho_0}\frac{\partial V_x}{\partial t}+V_z\frac{\partial V_x}{\partial z}\right) 
    +\frac{3\eta_0}{\rho_0}\frac{\partial^2}{\partial t\partial z}\left(\frac{b}{B_0}\left[\frac{b}{B_0}\frac{\partial V_x}{\partial z}+\frac{2}{3}\frac{\partial V_z}{\partial z}\right]\right) + O(\epsilon^{5/2}).
\end{equation}
Interpreting Eq.~(\ref{eq:nonlinear_wave_eq_raw}), the linear terms (on the left hand side) correspond to a wave equation with wave speed $v_A$. The leading nonlinear terms (those shown explicitly on the right hand side) include the leading-order effect of the anisotropic viscosity, which enters at the same order as the leading nonlinear terms that appear in perturbative nonlinear theory of ideal Alfv\'en waves.

Next, we eliminate $b$ and $\delta \rho$ from the $O(\epsilon^{3/2})$ nonlinear terms in Eq.~(\ref{eq:nonlinear_wave_eq_raw}).
Equations~(\ref{eq:dbdt}) and (\ref{eq:dVxdt}) are solved at linear order by the Alfv\'en wave relation
\begin{equation}
    \frac{b}{B_0} = \pm \frac{V_x}{v_A} + O(\epsilon^{3/2}).
\end{equation}
We choose the negative sign so waves travel in the positive $z$ direction, giving
\begin{equation}\label{eq:b_of_Vx}
    \frac{b}{B_0} = - \frac{V_x}{v_A} + O(\epsilon^{3/2}).
\end{equation}
The travelling wave behaviour of the linear solution together with assumption that the wave envelope changes over a distance controlled by the leading order nonlinear terms in Eq.~(\ref{eq:nonlinear_wave_eq_raw}) allows replacement
\begin{equation}\label{eq:wave_derivs_err}
    \frac{\partial}{\partial t } = -v_A \frac{\partial}{\partial z} + O(\epsilon).
\end{equation}

The density perturbation is governed by the mass continuity equation
\begin{equation}
    \frac{\partial \rho}{\partial t} + \nabla\cdot\left(\rho\mathbf{V}\right)=0,
\end{equation}
which gives for our shear Alfv\'en wave
\begin{equation}
    \frac{\partial \delta\rho}{\partial t} = -\rho_0 \frac{\partial V_z}{\partial z} + O(\epsilon^2).
\end{equation}
Then, using Eq.~(\ref{eq:wave_derivs_err}) and integrating,
\begin{equation}\label{eq:rho_of_Vz}
    \frac{\delta\rho}{\rho_0} = \frac{V_z}{v_A} + O(\epsilon^2).
\end{equation}
The constant of integration has been set to zero, for reasons discussed in \S~\ref{sec:nocera}.

Using these results, Eq.~(\ref{eq:nonlinear_wave_eq_raw}) becomes
\begin{equation}\label{eq:nonlinear_wave_eq_Vx_Vz}
    \left(\frac{\partial^2}{\partial t^2}-v_A^2\frac{\partial^2}{\partial z^2}\right)V_x
    = v_A \frac{\partial^2}{\partial z^2}\left(V_xV_z\right)
     + \frac{3\eta_0}{\rho_0}\frac{\partial^2}{\partial t \partial z}
    \left(\frac{V_x}{v_A}\left[\frac{V_x}{v_A}\frac{\partial V_x}{\partial z}-\frac{2}{3}\frac{\partial V_z}{\partial z}\right]\right)  + O(\epsilon^{5/2}).
\end{equation}

Now that the problem has been reduced to the two variables $V_x$ and $V_z$, it is convenient to make the $\epsilon$ dependence explicit by introducing dimensionless variables $v$ and $w$ defined by
\begin{equation}
    V_x(z,t) = \epsilon^{1/2}v_A v(z,t), \quad
    V_z(z,t) = \epsilon v_A w(z,t).
\end{equation}
Expressing Eq.~(\ref{eq:nonlinear_wave_eq_Vx_Vz}) in the dimensionless variables $v$ and $w$, and dropping the non-explicit higher order terms from the right hand side, we seek solutions to
\begin{equation}\label{eq:nonlinear_wave_eq_v_w}
    \left(\frac{\partial^2}{\partial t^2}-v_A^2\frac{\partial^2}{\partial z^2}\right)v
    = \epsilon\left(v_A^2\frac{\partial^2}{\partial z^2}(vw)+\frac{\eta_0}{\rho_0}\frac{\partial^2}{\partial t \partial z}
    \left(\frac{\partial v^3}{\partial z}-2v\frac{\partial w}{\partial z}\right)\right).
\end{equation}

\subsubsection{Multiple scale analysis}

Equation~(\ref{eq:nonlinear_wave_eq_v_w}) is now solved using multiple scale analysis \citep[e.g.][]{1978BenderOrszag}.
Applying this technique, one introduces a new variable $Z=\epsilon z$ that defines a long length scale, and the perturbation expansions
\begin{eqnarray}
    v(z,t) &=& v_0(z,Z,t) + \epsilon v_1(z,Z,t) + \ldots \label{eq:v_expanded} \\
    w(z,t) &=& w_0(z,Z,t) + \epsilon w_1(z,Z,t) + \ldots . \label{eq:w_expanded}
\end{eqnarray}
Derivatives are treated using the chain rule as though $z$ and $Z$ are independent variables and setting $\mathrm{d}Z/\mathrm{d}z=\epsilon$. Thus,
\begin{eqnarray}
    \frac{\partial v}{\partial z} &=& 
    \frac{\partial v_0}{\partial z}
    + \epsilon \left( \frac{\partial v_0}{\partial Z} + \frac{\partial v_1}{\partial z} \right) + O(\epsilon^2), \\
    \frac{\partial^2 v}{\partial z^2} &=& 
    \frac{\partial^2 v_0}{\partial z^2}
    + \epsilon \left( 2\frac{\partial^2 v_0}{\partial Z \partial z} + \frac{\partial^2 v_1}{\partial z^2} \right) + O(\epsilon^2),
\end{eqnarray}
with equivalent expressions for derivatives of $w$.
    
Substituting into Eq.~(\ref{eq:nonlinear_wave_eq_v_w}), collecting $\epsilon^0$ terms, and thus solving the homogeneous wave equation
\begin{equation}
    \left(\frac{\partial^2}{\partial t^2}-v_A^2\frac{\partial^2}{\partial z^2}\right)v_0
    = 0,
\end{equation}
obtains d'Alembert's solution
\begin{equation}
    v_0(z,Z,t) = f(z-v_A t,Z) + g(z+v_A t,Z).
\end{equation}
For forward propagating waves, the function $g$ is zero.

The corresponding $w_0$ is obtained by integrating the $z$-component of the momentum equation, Eq.~(\ref{eq:mom}), which gives
\begin{equation}\label{eq:Vz_relation_intermediate}
    \frac{V_z}{v_A} = \frac{1}{\rho_0 v_A^2}\left(\delta p +\frac{b^2}{2\mu_0}+\pi_{zz}\right) + O(\epsilon^2).
\end{equation}
From Eqs.~(\ref{eq:pi}) and (\ref{eq:W0}), we have
\begin{equation}\label{eq:pi_zz}
    -\pi_{zz}=2\eta_0\left(\frac{b}{B_0}\frac{\partial V_x}{\partial z}+\frac{2}{3}\frac{\partial V_z}{\partial z}\right) + O(\epsilon^{2}).
\end{equation}
A substitution for the pressure perturbation $\delta p$ is obtained by integrating the energy equation,
\begin{equation}\label{eq:energy}
    \frac{\partial p}{\partial t}+\mathbf{V}\cdot\nabla p + \gamma p \nabla\cdot\mathbf{V} = (\gamma-1)Q_{visc}.
\end{equation}
The viscous heating $Q_{visc}$ is of order $O(\epsilon^2)$, so integration gives the adiabatic relation
\begin{equation}\label{eq:p_of_Vz}
    \frac{\delta p}{p_0} = \gamma \frac{V_z}{v_A} + O(\epsilon^2).
\end{equation}
Alternatively, one can consider isothermal conditions using $\delta p / p_0 = V_z/v_A$ from the ideal gas law, which is recovered from Eq.~(\ref{eq:p_of_Vz}) by setting $\gamma=1$.

Using Eqs.~(\ref{eq:pi_zz}) and (\ref{eq:p_of_Vz}), and eliminating $b$ terms using Eq.~(\ref{eq:b_of_Vx}), Eq.~(\ref{eq:Vz_relation_intermediate}) can be expressed as
\begin{equation}\label{eq:w_v_pde}
    \left(1-\beta+\frac{4}{3}\frac{\eta_0}{\rho_0v_A}\frac{\partial}{\partial z}\right)w
    = \left(\frac{1}{2}+\frac{\eta_0}{\rho_0 v_A}\frac{\partial }{\partial z}\right)v^2,
\end{equation}
where $\beta=(c_s/v_A)^2$ and dropped terms are $O(\epsilon)$.
When $v$ and $w$ are expanded according to Eqs.~(\ref{eq:v_expanded}) and (\ref{eq:w_expanded}),
an equation identical to (\ref{eq:w_v_pde}) connects $w_0$ and $v_0$.

Inspecting Eq.~(\ref{eq:w_v_pde}), it is evident that obtaining $w$ for a known $v$ in general requires solving a first order linear partial differential equation. In the limit where the viscous terms can be neglected, the problem simplifies to the algebraic $w=v^2/2(1-\beta)$ relation used in Sec.~\ref{sec:compressible}. Similarly, when the viscous terms dominate, one obtains the $w=(3/4)v^2$ relation for viscously self-organised parallel flow discussed in Sec.~\ref{sec:compressible}. For the detailed solution in this section, we retain the complete set of forces that determine $V_z$, solving the full  Eq.~(\ref{eq:w_v_pde}).

Solution is facilitated by considering the special case where $v_0$ oscillates sinusoidally in time. For the rest of this derivation we therefore set
\begin{align}\label{eq:v0}
    v_0 &= A(Z)\mathrm{e}^{i \phi} + A^*(Z) \mathrm{e}^{-i \phi}, &
    \phi &= k_{||}(z-v_A t),
\end{align}
where $A(Z)\in\mathbb{C}$ and $*$ denotes the complex conjugate.
Representing $A$ in polar form,
\begin{equation}\label{eq:polar}
    A(Z) = R(Z)\mathrm{e}^{i\theta(Z)},
\end{equation}
Eq.~(\ref{eq:v0}) is equivalent to
\begin{equation}\label{eq:cosine}
    v_0(z,Z,t) = 2R(Z)\cos(k_{||}(z-v_A t)+\theta(Z)).
\end{equation}
From inspection, $2R(Z)$ is the local amplitude and $\theta(Z)$ is a phase shift. 
We have found the complex form in Eq.~(\ref{eq:v0}) more convenient to work with in the following.

We now solve for the corresponding $w_0$. Noting that
\begin{equation}
    v_0^2 = A^2\mathrm{e}^{2i\phi}+2|A|^2+{A^*}^2\mathrm{e}^{-2i\phi},
\end{equation}
where $|A|^2=AA^*$, we seek a solution of the form
\begin{equation}\label{eq:w0}
    w_0 = D\mathrm{e}^{2i\phi}+D^*\mathrm{e}^{-2i\phi}+\hat{w}_0.
\end{equation}
Substituting into Eq.~(\ref{eq:w_v_pde}), terms in $\mathrm{e}^0$ give
\begin{equation}\label{eq:w0_hat}
    \hat{w}_0 = \frac{|A|^2}{1-\beta},
\end{equation}
while terms in $\mathrm{e}^{2i\phi}$ and $\mathrm{e}^{-2i\phi}$ independently give
\begin{equation}\label{eq:alpha}
    D = \alpha A^2, \quad 
    \alpha = \frac{1}{2}\frac{(1+4ik_{||}\eta_0/\rho_0v_A)}{(1-\beta+(8/3)ik_{||}\eta_0/\rho_0v_A)}.
\end{equation}

The solution for $w_0$ can also be expressed without complex numbers. Making explicit the real and imaginary parts of $\alpha=\alpha_r+i\alpha_i$, we have the real constants
\begin{eqnarray}
    \alpha_r &=& \frac{1}{2}\frac{(1-\beta+(32/3)(\mathrm{Re})^{-2})}{((1-\beta)^2+(64/9)(\mathrm{Re})^{-2})}, \label{eq:alpha_r} \\
    \alpha_i &=& \frac{2}{3}\frac{(1-3\beta)(\mathrm{Re})^{-1}}{((1-\beta)^2+(64/9)(\mathrm{Re})^{-2})},
    \label{eq:alpha_i}
\end{eqnarray}
where $\mathrm{Re}=\rho_0 v_A / k_{||}\eta_0$ consistent with Eq.~(\ref{eq:Re}).
It is then easily shown that
\begin{equation}
     \frac{w_0}{2R(Z)^2}=
    \alpha_r\cos(2(k_{||}(z-v_At)+\theta(Z))) \\
    -\alpha_i\sin(2(k_{||}(z-v_At)+\theta(Z)))
    +\frac{1}{2(1-\beta)}.
\end{equation}

To deduce $R(Z)$ and $\theta(Z)$, we return to analysing Eq.~(\ref{eq:nonlinear_wave_eq_v_w}).
The $\epsilon^1$ terms in Eq.~(\ref{eq:nonlinear_wave_eq_v_w}) give the inhomogeneous partial differential equation
\begin{equation}\label{eq:v1}
    \left(\frac{\partial^2}{\partial t^2}-v_A^2\frac{\partial^2}{\partial z^2}\right)v_1
    =  
    2v_A^2\frac{\partial^2 v_0}{\partial Z \partial z}
    + v_A^2 \frac{\partial^2}{\partial z^2}\left(v_0w_0\right)
    + \frac{\eta_0}{\rho_0}\frac{\partial^2}{\partial t \partial z}
    \left(\frac{\partial v_0^3}{\partial z}-2v_0\frac{\partial w_0}{\partial z}\right).
\end{equation}
The $v_0$ and $w_0$ terms drive $v_1$, and the solution for $v_1$ will have a secular contribution (i.e. one or more terms that grow relative to corresponding solutions of the homogeneous equation) if terms on the right hand side resonate with the solution to the undriven wave equation. 
In the specific case where $v_0$ is given by Eq.~(\ref{eq:v0}), secular terms in the solution for $v_1$ will restrict the domain for which $v_0$ is a valid approximation if the right hand side of Eq.~(\ref{eq:v1}) contains $\mathrm{e}^{i\phi}$ or $\mathrm{e}^{-i\phi}$ terms. 
The central idea in multiple scale analysis is to solve for the $A(Z)$ that makes the resonance disappear, making $v_0$ a durable approximation for $v$.

Using Eqs.~(\ref{eq:v0}) and (\ref{eq:w0})--(\ref{eq:alpha}), $\mathrm{e}^{i\phi}$ terms vanish from the right hand side of Eq.~(\ref{eq:v1}) if and only if
\begin{equation}\label{eq:dAdz}
    \frac{1}{A}\frac{\mathrm{d}A}{\mathrm{d}Z} = 
    -\frac{k_{||}|A|^2}{2}\left[i\left(\alpha+\frac{1}{1-\beta}\right)
    +\frac{k_{||}\eta_0 }{\rho_0v_A}\left(3-4\alpha\right)
    \right].
\end{equation}
The same condition also removes the $\mathrm{e}^{-i\phi}$ terms.

Changing to polar form, Eq.~(\ref{eq:polar}) implies
\begin{equation}
    \frac{1}{A}\frac{\mathrm{d}A}{\mathrm{d}Z} = 
    \frac{1}{R}\frac{\mathrm{d}R}{\mathrm{d}Z} + i \frac{\mathrm{d}\theta}{\mathrm{d}Z}.
\end{equation}
Hence, the real and imaginary parts of Eq.~(\ref{eq:dAdz}) yield the real ordinary differential equations
\begin{align}
 \frac{\mathrm{d}R}{\mathrm{d}Z} &= -\frac{\kappa_1}{2} R^3, 
    \label{eq:ode_polar_R} \\
  \frac{\mathrm{d}\theta}{\mathrm{d}Z} &= \kappa_2R^2, 
    \label{eq:ode_polar_theta}
\end{align}
where 
\begin{align}
    \kappa_1 &= k_{||}\left(\frac{k_{||}\eta_0}{\rho_0v_A}\left(3-4\alpha_r\right)-\alpha_i\right), \label{eq:kappa_1_alphas} \\
    \kappa_2 &= -\frac{k_{||}}{2}\left(\alpha_r+\frac{1}{1-\beta}-\frac{k_{||}\eta_0}{\rho_0 v_A}4\alpha_i\right). \label{eq:kappa_2_alphas}
\end{align}
Eqs.~(\ref{eq:ode_polar_R}) and (\ref{eq:ode_polar_theta}) govern the local amplitude and phase drift of the Alfv\'en wave respectively (\textit{c.f.} Eq.~(\ref{eq:cosine})).

Our main interest is in $R(Z)$, which determines how the waves decay. 
Equation~(\ref{eq:ode_polar_R}) is a separable first order differential equation. The solution is
\begin{equation}\label{eq:R_sol}
    R(Z) = \frac{R(0)}{\sqrt{1+\kappa_1 R(0)^2 Z}}.
\end{equation}
For $\kappa_1>0$ the wave envelope decays non-exponentially, over a damping length that is inversely proportional to the square of the initial wave amplitude. 
Having obtained $R(Z)$, the solution for $\theta(Z)$ is obtained by directly integrating Eq.~(\ref{eq:ode_polar_theta}). 
Using Eq.~(\ref{eq:R_sol}),
\begin{equation}
    \theta(Z) = \theta(0)+\frac{\kappa_2}{\kappa_1}\ln\left|1+\kappa_1 R(0)^2 Z\right|.\label{eq:theta_sol}
\end{equation}

\subsubsection{Solution in original variables}

Having ensured corrections to $v\approx v_0$ remain of order $\epsilon\sim(b/B_0)^2\ll1$, the multiple scale analysis is concluded by using $v_0$ as the approximation for $v$. 
Returning to the original variables,
\begin{equation}\label{eq:Vx_sol}
    V_x(z,t) = \frac{a_0}{\sqrt{1+z/L_d}}
    \cos\left(
    k_{||}(z-v_A t)+(\kappa_2/\kappa_1)\ln\left|1+z/L_d\right| + \theta_0
    \right),
\end{equation}
where $a_0$ is the amplitude of $V_x(0,t)$, $\theta_0$ sets the initial phase of the wave (at $z=0$, $t=0$) and
\begin{equation}\label{eq:Ld_kappa1}
    L_d = \frac{4}{\kappa_1 (a_0/v_A)^2}
\end{equation}
is the decay length.
Using Eqs.~(\ref{eq:alpha_r}), (\ref{eq:alpha_i}) and (\ref{eq:kappa_1_alphas}),
\begin{equation}
    \kappa_1 = \frac{k_{||}}{3\mathrm{Re}}\frac{(1-3\beta)^2}{(1-\beta)^2+(64/9)(\mathrm{Re})^{-2}},
\end{equation}
where $\mathrm{Re}$ is the Reynolds number for the wave, as defined by Eq.~(\ref{eq:Re}).
Thus,
\begin{equation}\label{eq:Ld_full}
    L_d = \frac{12 \mathrm{Re} }{k_{||}(a_0/v_A)^2}\frac{(1-\beta)^2+(64/9)(\mathrm{Re})^{-2}}{(1-3\beta)^2}.
\end{equation}

If one neglects the $\mathrm{Re}^{-2}$ term in the numerator of Eq.~(\ref{eq:Ld_full}), then $L_d$ agrees exactly with the formula in Eq.~(\ref{eq:damping:length}) that we derived from energy principles.
The formula for $L_d$ in Eq.~(\ref{eq:Ld_full}) is more general since it was derived without direct assumptions about the value of $\mathrm{Re}=k_{||}\eta_0/\rho_0 v_A$, although the multiple scale analysis requires that the combination of parameters $k_{||}$, $a_0^2$ and $\mathrm{Re}$ are such that waves damp over a significantly longer scale than the wavelength.

\subsection{Non-exponential decay and interpretation of damping length}

As a general principle, the Alfv\'en wave energy density $E_w = \rho V_x^2$ decays more rapidly than the perturbation $V_x$, due to the quadratic power. For exponential decay this is reflected in a factor two difference in the respective $\mathrm{e}^{-1}$ decay lengths. For the non-exponential decay produced by nonlinear viscous damping, the situation is handled differently. The same $L_d$ describes $V_x$ and $E_w$, however they have different functional forms.
The velocity amplitude decays as $(1+z/L_d)^{-1/2}$ (see Eq.~(\ref{eq:Vx_sol})), while the wave energy density decays as $(1+z/L_d)^{-1}$. Therefore, over a distance $L_d$, the velocity amplitude reduces by a factor $\sqrt{2}$ and the energy density halves.

\subsection{Inclusion of thermal conduction}
The multiple scale analysis can also be modified to include explicit thermal conduction. Since thermal conduction is highly anisotropic, we include the parallel thermal conduction, setting the heat flow vector to 
\begin{equation}
    \mathbf{q} = -\mathcal{K}_{||} (\mathbf{h}\cdot\nabla T) \mathbf{h},
\end{equation}
where $\mathcal{K}_{||}$ is the coefficient of parallel thermal conduction, and temperature $T = p/\rho \mathcal{R}$ where $\mathcal{R}$ is the gas constant.
The energy equation with heat flow is
\begin{equation}
    \frac{\partial p}{\partial t}+\mathbf{V}\cdot\nabla p + \gamma p \nabla\cdot\mathbf{V} = (\gamma-1)(Q_{visc}-\nabla\cdot\mathbf{q}),
\end{equation}
which replaces Eq.~(\ref{eq:energy}).

It follows that $\delta p / p_0$ is related to $v_z/v_A$ by the partial differential equation
\begin{equation}\label{eq:p_of_Vz_conduction}
    \left(1+\Lambda\frac{\partial }{\partial z}\right)\frac{\delta p}{p_0}
    = 
    \left(\gamma+\Lambda\frac{\partial }{\partial z}\right)\frac{V_z}{v_A}+O(\epsilon^2).
\end{equation}
where 
\begin{equation}
    \Lambda=\frac{(\gamma-1)\mathcal{K}_{||}}{\rho_0 \mathcal{R} v_A}
\end{equation} 
is a conductive length scale.
In the limit of weak thermal conduction, $\Lambda\to0$ gives $\delta p / p_0 = \gamma V_z/v_A$, recovering the adiabatic case treated above. Similarly, for strong thermal conduction, $\Lambda\to\infty$ gives $\delta p / p_0 = V_z/v_A$, recovering the isothermal case.

Introducing a dimensionless pressure variable $c$ defined by $\delta p = \epsilon p_0 c(z,t)$, expanding $c(z,t)=c_0(z,Z,t)+\epsilon c_1(z,Z,t)+\ldots$ and setting $c_0=C\mathrm{e}^{2i\phi}+C^*\mathrm{e}^{2i\phi}+\hat{c}_0$, terms in $\mathrm{e}^0$ in Eq.~(\ref{eq:p_of_Vz_conduction}) yield $\hat{c}_0=\gamma \hat{w}_0$, and
terms in $\mathrm{e}^{2i\phi}$ yield $C=\Gamma D$, where 
\begin{equation}
    \Gamma = \frac{\gamma+2ik_{||}\Lambda}{1+2ik_{||}\Lambda}.
\end{equation}

Solving further, an equation $D=\alpha A^2$ analogous to Eq.~(\ref{eq:alpha}) is obtained but with $\beta$ replaced by the complex-valued $\Gamma p_0 / \rho_0 v_A^2$ in the formula for $\alpha$. Meanwhile, (\ref{eq:w0_hat}) and (\ref{eq:dAdz}) are unchanged, retaining the real-valued $\beta=\gamma p_0/\rho_0 v_A^2$.
The wave amplitude is therefore governed by results identical to Eqs.~(\ref{eq:ode_polar_R}) and (\ref{eq:kappa_1_alphas}), with the aforementioned change in the definition of $\alpha$.

\section{Discussion}\label{sec:disc}

\subsection{Optimum damping}\label{sec:optimum}
Inspecting Eq.~(\ref{eq:Ld_full}), the formula for $L_d k_{||}$ has a minimum with respect to the Alfv\'enic Reynolds number at $\mathrm{Re}=8/(3|1-\beta|)$. Thus, shear Alfv\'en waves with $k_{||}=3\rho_0 v_A |1-\beta|/3\eta_0$ are damped in the fewest number of wavelengths, which we refer to as optimum damping. The optimally damped waves have
\begin{equation}\label{eq:Ld_optimum}
    \frac{L_d}{\lambda_{||}} = \frac{32}{\pi} \frac{|1-\beta|}{(a_0/v_A)^2(1-3\beta)^2}.
\end{equation}

When $\beta\ll 1$, the right hand side of Eq.~(\ref{eq:Ld_optimum}) is approximately ten divided by the square of the normalised wave amplitude. Hence, while nonlinear viscous damping can in principle damp Alfv\'en waves in a small number of wavelengths, this requires large amplitudes $a/v_A\sim 1$, or $\beta\sim 1$. For a more typically encountered amplitudes $a/v_A\sim 10^{-1}$ and low $\beta$, one finds $L_d/\lambda_{||}\gtrsim1000$, making nonlinear viscous damping negligible for many coronal situations.

\subsection{Viscous self-organisation}
The suppression of nonlinear viscous damping for small $\mathrm{Re}$ (highly viscous plasma) does not mean that viscous effects are unimportant in this regime. On the contrary, nonlinear damping is suppressed for small $\mathrm{R_e}$ because viscous forces organise the parallel flow associated with the Alfv\'en wave to approach the relationship $V_z/v_A = (3/4)(V_x/v_A)^2$. This modification of the parallel flow plays a crucial role in avoiding significant nonlinear damping in highly viscous plasma, when modelled using Braginskii MHD.

\subsection{Validity constraints}\label{sec:validity}
Throughout this paper, we have assumed that $\beta\neq1$ to avoid resonant wave coupling. This condition holds throughout most of the corona, so it is appropriate for our primary applications. Additionally, the multiple scale analysis in \S~\ref{sec:damping} uses $\epsilon\sim(V_x/v_A)^2\sim(b/B_0)^2$ as a small parameter, one consequence of which is that the nonlinearly damping occurs over a distance considerably greater than the parallel wavelength. As noted in \S~\ref{sec:params}, transverse coronal waves are observed in open-field regions with $\epsilon\sim10^{-2}$, making weakly nonlinear theory appropriate for such situations. Obtaining nonlinear viscous solutions in the resonant and strongly nonlinear regimes nonetheless remain interesting future challenges for plasma theory.

Applicability of this paper's results to physical problems is also constrained to conditions under which Braginskii MHD can be rigorously applied. As discussed in \S~\ref{sec:bg}, the traditional derivation of Braginskii MHD assumes that the collisional mean free path is less than the macroscopic scales. Comparing the mean free path parallel to the magnetic field to the parallel wavelength, this condition can be given as $k_{||} \lambda_{mfp}  < 1$, where $\lambda_{mfp}=v_{Ti}\tau_i$, $v_{Ti}=\sqrt{k_B T/m_i}$ and $\tau_i$ is the ion collision time. Using the formula \citep{1958Braginskii, 1965Braginskii,1985Hollweg}
\begin{equation}
    \eta_0 = 0.96 n k_B T \tau_i,
\end{equation}
and the definition of $\mathrm{Re}$ in Eq.~(\ref{eq:Re}), one can show that
\begin{equation}\label{eq:validity}
    k_{||} \lambda_{mfp} < 1 \ \Leftrightarrow \ \beta^{1/2}\mathrm{Re} = \frac{\rho c_s}{k_{||}\eta_0} > 1.
\end{equation}
In other words, Braginskii MHD requires that the Reynolds number based on the sound speed is greater than unity.
One should therefore be cautious about applying small Alfv\'enic Reynolds number results such as viscous self-organisation to real low-$\beta$ plasmas.

\subsection{Formulas for applications}

For applications to real plasmas, the following formulas are convenient.
In cases where the parallel viscosity coefficient is determined by Coulomb collisions,
\begin{equation}\label{eq:cor:eta0}
    \eta_0 = 0.96 n k_B T \tau_i = \frac{22}{\lambda_C}\times10^{-17}T^{5/2},
\end{equation}
where this formula is stated in S.I. units with $T$ in kelvin, and $\lambda_C$ is the Coulomb logarithm \citep[e.g.][]{1985Hollweg}.
The Reynolds number defined in Eq.~(\ref{eq:Re}) can then be expressed as 
\begin{equation}\label{eq:Re_physical}
    \mathrm{Re}=5.8\times10^{20} \lambda_CB^2 f^{-1}T^{-5/2},
\end{equation}
also in S.I. units, where $f=v_A k_{||}/2\pi$ is the wave frequency.
This formula makes explicit the dependences on frequency, magnetic field strength and temperature. The Alfv\'enic Reynolds number is smallest when the plasma has high temperature and low magnetic field strength, and for higher frequency waves.
Finally, we express the damping length in Eq.~(\ref{eq:Ld_full}) as a function of frequency and the mean square velocity $\left<V_x^2\right>=a_0^2/2$, which gives
\begin{equation}\label{eq:Ld_freq}
    L_d = \frac{3}{\pi} \frac{v_A^3 \mathrm{Re} }{f\left<V_x^2\right>} \frac{(1-\beta)^2+(64/9)(\mathrm{Re})^{-2}}{(1-3\beta)^2}.
\end{equation}

\subsection{Waves in a coronal open-field region}

Outgoing transverse waves in the magnetically open solar corona contain sufficient energy to heat the open corona and accelerate the fast solar wind \citep{2011McIntosh,2015Morton}, and they are observed to damp significantly within a solar radius above the Sun's surface \citep{2012BemporadAbbo,2012Hahn,2013HahnSavin,2022Hahn}. Heating at these altitudes is also thought to be important for producing the observed rapid acceleration of the fast solar wind \citep{1995Habal,1995McKenzie}. The problem of how the outgoing waves damp has not been conclusively solved, although one leading hypothesis is turbulent cascade driven by interactions with downgoing Alfv\'en waves \citep{1986Hollweg,1992HeyvaertsPriest,1999Matthaeus,2007Cranmer,2010Verdini,2018Mikic} produced either by reflection from density inhomogeneities \citep{2016VanBallegooijen,2022Pascoe} or by parametric decay instability \citep{1963GaleevOraevskii,1978Derby,1978Goldstein,2019Shoda,2022Hahn}.

Here, we demonstrate that Braginskii viscosity does not cause significant damping of Alfv\'en waves at the altitudes at which the traditional derivation of Braginskii MHD holds. For concreteness, we consider the Sun’s northern polar open-field region on 27 March 2012, using observational values reported by \citet{2015Morton}. Enhanced wave power was present around $f = 5 \mbox{ mHz}$, which suggests Alfv\'enic waves produced by p-modes \citep{2019Morton}. We will calculate damping lengths for this frequency, noting that $\mathrm{Re}$ and $L_d$ depend on $f$, with $L_d\sim f^{-2}$ in the limit of high $\mathrm{Re}$. \citet{2015Morton} inferred that the Alfv\'en speed was nearly constant with $v_A = 400 \mbox{ km s}^{-1}$ on their domain of $r = 1.05$ to $1.20 R_\sun$. For temperature, we set $T = 1.6 \times 10^6 \mbox{ K}$, the formation temperature of the Fe XIII lines used by the CoMP instrument, which implies the proton thermal speed $V_{Ti}=\sqrt{k_B T / m_i}$ is $115 \mbox{ km s}^{-1}$. Hence, in for an isothermal equation of state $\beta = 0.083$ and $\beta^{1/2}=0.29$. For the wave velocity amplitude, \citet{2015Morton} recommended that the reported non-thermal line width should be used, which varies with altitude. 

Starting with lowest altitude observed by \citet{2015Morton}, $r=1.05 R_\sun$, we set $n=10^{14} \mbox{ m}^{-3}$, $B=2\times10^{-4} \mbox{ T}$ and take the rms value of $V_x$ as $35 \mbox{ km s}^{-1}$. We therefore find $\lambda_C = 19$ and $\mathrm{Re}=28$. Since $\beta^{1/2}\mathrm{Re}=8>1$, Braginksii MHD applies and we evaluate $L_d=4.2\times10^8 \mbox{ km} \equiv 600 R_\sun$. 

At $r=1.20$, we set $n=10^{13} \mbox{ m}^{-3}$, $B=6\times10^{-5} \mbox{ T}$ and take the rms value of $V_x$ as $50 \mbox{ km s}^{-1}$. The observed parameters therefore give $\lambda_C = 21$ and $\mathrm{Re}=2.7$. Since $\beta^{1/2}\mathrm{Re}=0.8\approx1$, this altitude is close to the maximum at which the assumptions by which Braginskii MHD is traditionally derived remains valid (for this particular open field region, and assuming Eq.~(\ref{eq:cor:eta0})). Evaluating the damping length here returns $L_d=4.2\times10^7\mbox{ km}\equiv61 R_\sun$. 

We conclude that Braginskii viscosity does not cause significant wave damping below $r=1.2 R_\sun$, which is consistent with observational results that Alfv\'enic wave amplitudes in coronal holes follow ideal WKB scaling out to around this altitude \citep{2005Cranmer,2013HahnSavin}. 

Between the altitudes we have examined, $L_d$ reduces by an order of magnitude. If one were to extrapolate using high $\mathrm{Re}$ or incompressible results, it would appear that viscous damping becomes important near the altitudes at which the waves are observed to damp. We are cautious about making such an assertion for two reasons. First, as discussed in \S~\ref{sec:optimum}, our results show that for $\mathrm{Re}<8/(3|1-\beta))$ the damping length in a Braginskii MHD model increases again as the field-aligned flow self organises to supress viscous damping. Secondly, as the plasma becomes increasingly collisionless ($\beta^{1/2}\mathrm{Re}<1$) the traditional derivation of Bragniskii MHD falters.

Intriguingly, it may be significant that the onset of wave damping broadly coincides with the altitude at which Braginskii MHD can no longer be confidently applied if one invokes the $\eta_0$ expression for Coulomb collisions given in Eq.~(\ref{eq:cor:eta0}). This correspondence is suggestive that the wave damping observed in coronal holes may involve collisionless and heat flow effects not found in the most common fluid models.

\subsection{Future work}
The present types of analyses should be extended in future to other types of propagating transverse MHD waves. The nonlinear longitudinal flow that accompanies propagating torsional Alfv\'en waves differs from its counterpart for propagating shear Alfv\'en waves \citep{2011VasheghaniFarahani} and it will be of interest to investigate how this difference affects the nonlinear viscous damping. It is similarly desirable to determine how nonlinear viscosity affects propagating kink waves \citep{1983EdwinRoberts}.

For propagating shear Alfv\'en waves, viscous damping appears most promising near the $c_s=v_A$ singularity, which must be treated using different methods to those used in this paper. The solar wind frequently has $\beta\sim1$, while $\beta=1$ occurs in the lower solar atmosphere and in the vicinity of coronal nulls points. Hence, this case is of considerable physical interest. 
One challenge for application to magnetic nulls is that the magnetic field unit vector $\mathbf{h}=\mathbf{B}/|\mathbf{B}|$ is not defined at the null itself, so one must be careful to evaluate the Braginskii stress tensor using appropriate calculations, e.g. see recent discussion by \citet{2017MacTaggart}.

A further challenge is to develop a theory of nonlinear viscous damping applicable to strongly nonlinear waves with amplitudes $b\sim B_0$ and greater. The results of the multiple-scale analysis in \S~\ref{sec:damping} are rigorous only for the weakly nonlinear case, in which $\epsilon\sim (b/B_0)^2$ can be treated as a small parameter and it is assumed that the damping length is significantly longer than the wavelength. Strongly nonlinear Alfv\'en waves with $b\sim B_0$ are a feature of the solar wind, and while the low collisionality of the solar wind means that Braginskii MHD may not be an appropriate framework for that application, extending the current work to strongly nonlinear waves remains an interesting problem.

There is a diverse collection of MHD wave problems beyond wave damping for which viscous effects are likely to be significant. Prime among these are nonlinear phenomena involving Alfv\'en waves, for which the nonlinear viscosity tensor enters the equations at the same order as the effect of interest. For example, standing Alfv\'en waves drive significantly stronger field-aligned flows than occur for propagating waves because standing Alfv\'en waves create inhomogeneous time-averaged magnetic pressure. There could also be significant value in investigating how viscosity modifies wave interactions, including Alfv\'en wave collisions and parametric decay instability \citep{1963GaleevOraevskii,1978Derby,1978Goldstein}, which are central to leading hypotheses of wave heating in the magnetically open solar corona.

Finally, we point to the continuing need for basic plasma physics research to provide increasingly rigorous derivation and validation of the appropriate fluid equations for weakly collisional and collisionless plasma, in the face of the closure problem summarised in \S~\ref{sec:bg}. As discussed in \S~\ref{sec:bg} and \ref{sec:validity}, Braginskii MHD breaks down at higher altitudes in the corona as the plasma becomes increasingly collisionless (see Eqs.~(\ref{eq:validity}) and (\ref{eq:Re_physical})). The CGL double-adiabatic equations and other models that evolve the stress tensor may provide a more suitable framework in these conditions. \citet{2019HunanaPart1,2019HunanaPart2,2022Hunana} provide recent discussions of such models and their limitations. Alternatively, it may be necessary for the solar waves community to more widely adopt non-fluid plasma models. However, tractability of kinetic models remains a limiting factor, especially in light of the large separations between kinetic and macroscopic scales that are characteristic of the Sun's corona. Eloquent comments on these matters can be found in \citet{1996Montgomery}.
 
\section{Conclusions}\label{sec:conc}
This paper has investigated the properties of propagating shear Alfv\'en waves subject to the nonlinear effects of the Braginskii viscous stress tensor. The main points are as follows:

\begin{enumerate}
    \item For many plasma environments, including the low-altitude solar corona, Braginskii MHD provides a more accurate description of plasma than classical MHD does, by rigorously treating the stress tensor and thermal conduction. Stress tensor effects nonetheless remain relatively unexplored for many solar MHD phenomena.
    
    \item The dominant viscous effects for propagating shear Alfv\'en waves are nonlinear in the wave amplitude and occur through the ``parallel'' viscosity coefficient, $\eta_0$. Theoretical results based on linearizing the stress tensor with respect to the wave amplitude are only valid for amplitudes satisfying $(b/B_0)^2 \ll (\Omega_i\tau_i)^{-2}$. Such waves would be energetically insignificant under normal coronal conditions, hence nonlinear treatment is required.

    \item Compressibility and pressure affect the nonlinear field-aligned flow associated with shear Alfv\'en waves, hence they impact nonlinear wave damping. Both must be included to produce accurate coronal results.
    
    \item Braginskii viscosity damps propagating shear Alfv\'en waves nonlinearly, such that the primary wave fields $b$ and $V_x$ decay as $(1+z/L_d)^{-1/2}$, where the decay length 
    $$L_d = \frac{12 \mathrm{Re}}{k_{||}(a_0/v_A)^2}\frac{(1-\beta)^2+(64/9)(\mathrm{Re})^{-2}}{(1-3\beta)^2}.$$ 
    Here, $a_0$ is the initial velocity amplitude of the wave, $\beta=(c_s/v_A)^2$ and $\mathrm{Re}=\rho v_A/k_{||}\eta_0$ is the Alfv\'enic Reynolds number of the wave. The energy density decays as $(1+z/L_d)^{-1}$.
    
    \item Optimal damping (the minimum normalised damping length $k_{||} L_d$) is obtained when $\mathrm{Re}=8/(3|1-\beta|)$. For low $\beta$ plasma and $(a_0/v_A)\lesssim 10^{-1}$, one finds $L_d/\lambda_{||}\gtrsim1000$, indicating that nonlinear viscous damping is negligible for many coronal situations. 
    
    \item The asymptotic behaviour that $L_d\to\infty$ in the highly viscous regime $\mathrm{Re}\to0$ is attributed to  self-organisation of the parallel flow by viscous forces such that $V_z/v_A\approx (3/4)(V_x/v_A)^2$, which suppresses dissipation.
    
    \item Applicability of the Braginskii MHD solutions to real plasmas is constrained by the traditional derivation of Braginskii MHD assuming that $k_{||} \lambda_{mfp} < 1$ which is equivalent to $\beta^{1/2}\mathrm{Re} = \rho c_s/k_{||}\eta_0 > 1$. In other words, Braginskii MHD requires that the Reynolds number based on the sound speed is greater than unity. We therefore recommend that only the damping results for large Alfv\'enic Reynolds number should be applied to real coronal plasma, using the simplified formula $L_d = 12 \mathrm{Re}(1-\beta)^2/(k_{||}(a_0/v_A)^2(1-3\beta)^2))$ that has been derived in this paper by two different techniques: energy principles and multiple scale analysis.

    \item Application to transverse waves observed in a polar open-field region concludes that nonlinear Braginskii viscosity does not cause significant damping of the waves at the altitudes at which the assumptions by which Braginskii MHD is traditionally derived remain valid ($r\lesssim 1.2 R_\sun$ for the considered region and wave properties). Intriguingly, the observed onset of wave damping broadly coincides with the altitude at which Braginskii MHD can no longer be confidently applied if one invokes the $\eta_0$ expression for Coulomb collisions given in Eq.~(\ref{eq:cor:eta0}).
    
\end{enumerate}

\begin{acknowledgments}
This work was prompted by and benefited from conversations with Paola Testa, Bart De Pontieu, Vanessa Polito, Graham Kerr, Mark Cheung, Wei Liu, David Graham, Joel Allred, Mats Carlsson, Iain Hannah and Fabio Reale during a research visit to LMSAL funded by ESA's support for the IRIS mission (August 2018) and meetings of International Space Science Institute (Bern) International Team 355 (November 2018). I am grateful to Peter Cargill, Andrew Wright, Bart De Pontieu and Paola Testa for comments on an early draft (June 2019), and Declan Diver for encouragment to explore connections with pressure anisotropy. I thank the reviewer for considered and constructive suggestions, and several unnamed individuals for comments that also improved the manuscript. The research made use of the NRL Plasma Formulary, NASA's ADS Bibliographic Services and the British Library On Demand service with assistance from librarians at the University of Dundee.
\end{acknowledgments}

\bibliography{WaveHeating}{}
\bibliographystyle{aasjournal}

\end{document}